\documentclass[runningheads]{llncs}

\usepackage[english]{babel}
\usepackage{subcaption}
\usepackage{url}
\usepackage{colortbl}
\usepackage{amssymb}
\usepackage{stmaryrd}
\usepackage{amsmath}
\usepackage{mathrsfs}
\usepackage{amssymb}
\usepackage[ruled,vlined,boxed,linesnumbered]{algorithm2e}
\usepackage[left]{lineno}
\usepackage{xfrac}
\usepackage{nicefrac}
\usepackage[textsize=scriptsize,backgroundcolor=yellow!40,obeyDraft]{todonotes}
\usepackage[nonumberlist,acronym,sanitize=none]{glossaries}
\glsdisablehyper
\usepackage{comment}
\usepackage[pdftex, colorlinks=true, hyperfootnotes=true, hyperindex=true,
plainpages=false, pagebackref=false, pdfpagelabels=true, pdfstartview=FitH,
linkcolor=blue, citecolor=blue, urlcolor=blue,
bookmarks, bookmarksopen, bookmarksdepth=3]{hyperref}
\usepackage[capitalise,nameinlink]{cleveref}
\crefname{algocf}{Alg.}{Algs.}
\Crefname{algocf}{Algorithm}{Algorithms}
\crefname{section}{Sec.}{Secs.}
\Crefname{section}{Section}{Sections}
\usepackage{paralist}
\usepackage{multicol}
\usepackage{booktabs}
\usepackage{enumitem}
\usepackage{multirow}
\usepackage{rotating}
\usepackage{etaremune}
\usepackage{marginnote}
\usepackage{mathtools}
\usepackage{adjustbox}
\usepackage{ifthen}
\usepackage[normalem]{ulem}
\usepackage{lineno}
\usepackage{soul}
\usepackage{floatrow}
\usepackage{listings}
\crefname{lstlisting}{Listing}{Listings}
\usepackage{lipsum}
\usepackage{makecell}
\usepackage{diagbox}
\usepackage[scientific-notation=false,group-separator={,}]{siunitx}
\usepackage{microtype}
\usepackage[bottom]{footmisc}
\usepackage{pgfplots}
\pgfplotsset{compat=1.10}
\usepackage{tikz}
\usetikzlibrary{automata,
	arrows,
	calc,
	chains,
	backgrounds,
	datavisualization,
	datavisualization.formats.functions,
	decorations.pathmorphing,
	decorations.pathreplacing,
	decorations.shapes,
	graphs,
	matrix,
	patterns,
	positioning,
	shapes.geometric,
	shapes,
	shapes.geometric,
	shapes.symbols,
	trees
}
\let\llncssubparagraph\subparagraph
\let\subparagraph\paragraph
\usepackage{titlesec}
\let\subparagraph\llncssubparagraph
\titlespacing{\subsubsection}{0pt}{*0.5}{*0.5}

\input{addon/commands}
\newacronym[\glslongpluralkey={Business Processes}]{bp}{BP}{Business Process}
\newacronym{wf}{WF}{workflow}
\newacronym{bpi}{BPI}{Business Process Intelligence}
\newacronym{bpm}{BPM}{Business Process Management}
\newacronym{bpms}{BPMS}{Business Process Management System}
\newacronym{bpmn}{BPMN}{Business Process Model and Notation}
\newacronym{soa}{SOA}{Service-Oriented Architecture}
\newacronym{kpi}{KPI}{Key Performance Indicator}
\newacronym{wfms}{WfMS}{Workflow Management System}
\newacronym{pn}{PN}{Petri net}
\newacronym{cpn}{CPN}{colored Petri net}
\newacronym{xes}{XES}{eXtensible Event Stream}
\newacronym{po}{PO}{Partial Order}
\newacronym{tl}{TL}{Temporal Logic}  
\newacronym{ltl}{LTL}{Linear Temporal Logic}
\newacronym{fol}{FOL}{First Order Logic}
\newacronym{ltlf}{LTL$_f$}{Linear Temporal Logic on Finite Traces}
\newacronym{mso}{MSO}{Monadic Second Order Logic}
\newacronym{rex}{RE}{regular expression}
\def\Autom {\ensuremath{A}}
\newacronym[symbol=\Autom,longplural={finite state automata}]{fsa}{FSA}{finite state automaton}
\newacronym[symbol=\Autom,longplural={deterministic finite automata}]{dfa}{DFA}{deterministic finite automaton}
\newacronym[symbol=\Autom,longplural={deterministic finite state automata}]{dfs}{DFS}{deterministic finite state automaton}
\newacronym[symbol=\Autom,longplural={nondeterministic finite automata}]{nfa}{NFA}{nondeterministic finite automaton}
\def\Declare {\textsc{Declare}}
\newglossaryentry{declare}{%
	name={\Declare},description={a declarative process modelling language and notation}}
\newglossaryentry{task}{%
	name={task},description={the non-divisible, elementary activity}}

\newcommand{\taskize}[1] {\ensuremath{\scalebox{0.85}{\normalfont\textsf{#1}}}}
\def\taska {\taskize{a}}
\def\taskb {\taskize{b}}
\def\taskc {\taskize{c}}
\def\taskd {\taskize{d}}

\def\taskt {\taskize{t}}

\def\taskv {\taskize{v}}

\newglossaryentry{promod}{%
	name={process model},description={the model of a process}
}
\def\DeclaModel {\ensuremath{\mathcal{M}}}
\newglossaryentry{declamodel}{%
	name={declarative \glsentrytext{promod}},description={\glsentrydesc{promod}, expressed by means of constraints},
	symbol={\DeclaModel}
}

\newglossaryentry{mindeclamodel}{%
	name={discovered \glsentrytext{declamodel}},description={\glsentrydesc{declamodel}, discovered from an \glsentrytext{evtlog}},
	symbol={\DeclaModel}
}
\newglossaryentry{minerful}{%
	name={MINERful},description={the declarative process discovery algorithm \glsentrytext{minerful}}}
\newacronym{mf}{Mf}{\gls{minerful}}
\newglossaryentry{minerfulVac}{%
	name={MINERful Vacuity Checker},description={\glsentrytext{minerful}} algorithm with semantical vacuity detection}
\newacronym{mfv}{Mf-Vchk}{\gls{minerfulVac}}
\newacronym{dmm}{DMM}{Declare Maps Miner}
\newglossaryentry{decmapmin}{%
	name={Declare Maps Miner},description={the declarative process discovery algorithm \glsentrytext{decmapmin}}}
\newacronym{dmm2}{DM2}{Declare Miner 2}
\newglossaryentry{decmapmin2}{%
	name={Declare Miner 2},description={improvement of \glsentrytext{decmapmin} algorithm}}
\newglossaryentry{janus}{%
	name={Janus},description={the declarative process discovery algorithm \glsentrytext{janus}}}
\def\LogAlph {\ensuremath{\Sigma}}

\newglossaryentry{logalph}{
	name={log alphabet},description={the process alphabet, as reflected in a log},%
	symbol={\LogAlph}}
\def\Evt {\ensuremath{e}}
\newglossaryentry{evt}{
	name={event},description={a record of an instantaneous fact during the process enactment},%
	symbol={\Evt}}
\def\EvtTrace {\ensuremath{t}}

\newglossaryentry{evttrace}{
	name={trace},description={a sequence of \glsplural{evt}},%
	symbol={\EvtTrace}}
\def\EvtLog {\ensuremath{L}}

\newglossaryentry{evtlog}{
	name={event log},description={a collection of \glstext{evttrace}s},%
	symbol={\EvtLog}}
\def\Subsum {\ensuremath{\sqsubseteq}}

\newglossaryentry{subsum}{%
	name={subsumption},description={is subsumed by},%
	symbol={\Subsum}}

\newglossaryentry{relaxop}{%
	name={relaxation},description={relaxation operator, climbing the \glsentrytext{subsum} hierarchy}}

\newglossaryentry{acti}{%
	name={activation},description={the activation of a constraint}}
\newglossaryentry{target}{%
	name={target},description={target}}
\def\Cns {\ensuremath{C}}
\newglossaryentry{con}{%
	name={constraint},description={a temporal business rule},
	symbol={\Cns}
}

\newglossaryentry{welldef}{%
	name={well-defined},description={of \glsentrytext{con}s for which a finite non-empty trace exists that complies with them}
}
\newglossaryentry{cnspar}{%
	name={parameter},description={a parameter of a \glsentrytext{con}},
}
\newglossaryentry{cnsarity}{%
	name={arity},description={number of parameters of a \glsentrytext{con}},
}
\newglossaryentry{exi}{
	name={existence},
	description={constrains single activities}
}
\newglossaryentry{exicon}{
	name={\glsentrytext{exi} \glsentrytext{con}},
	description={constrains single activities}
}
\newglossaryentry{posicon}{
	name={position \glsentrytext{con}},
	description={constrains the position of activities}
}
\newglossaryentry{cardicon}{
	name={cardinality \glsentrytext{con}},
	description={limits the number of activities}
}
\newglossaryentry{rela}{
	name={relation},
	description={constraint on pairs of activities}
}
\newglossaryentry{relacon}{
	name={\glsentrytext{rela} \glsentrytext{con}},
	description={constraint on pairs of activities}
}
\newglossaryentry{unirelacon}{
	name={unidirectional \glsentrytext{relacon}},
	description={constraint on pairs of activities, out of which one is the activation, as the other is the target}
}
\newglossaryentry{unifwrelacon}{
	name={\glsentrytext{fw}-\glsentrytext{unirelacon}},
	description={constraint on pairs of activities, having the first parameter as the activation, and the second one as the target}
}
\def\FwCns {\ensuremath{\mathit{fw}}}
\newglossaryentry{fw}{
	name={forward},
	description={forward constraint},
	symbol={\FwCns}
}

\newglossaryentry{unibwrelacon}{
	name={\glsentrytext{bw}-\glsentrytext{unirelacon}},
	description={constraint on pairs of activities, having the second parameter as the activation, and the first one as the target}
}
\def\BwCns {\ensuremath{\mathit{bw}}}
\newglossaryentry{bw}{
	name={backward},
	description={backward constraint},
	symbol={\BwCns}
}

\newglossaryentry{corelacon}{
	name={coupling \glsentrytext{con}},
	description={constraint based on pairs of relation constraints}
}
\newglossaryentry{nega}{
	name={negative},
	description={of a constraint, that negates a coupling relation constraint}
}
\newglossaryentry{negacon}{
	name={\glsentrytext{nega} \glsentrytext{con}},
	description={constraint negating a coupling relation constraint}
}
\def\CnsTemp {\ensuremath{\mathcal{C}}}
\newglossaryentry{cnstemp}{%
	name={template},description={the template of a \glsentrydesc{con}},
	symbol={\CnsTemp}}
\def\CnsTempPrime {\ensuremath{\CnsTemp'}}
\def\CnsTempSecond {\ensuremath{\CnsTemp''}}
\newcommand{\CnsTempFunc}[2] {\ensuremath{\CnsTemp(#1\ifthenelse{\equal{#2}{}}{}{,#2})}}
\newcommand{\CnsTempFuncPrime}[2] {\ensuremath{\CnsTempPrime(#1\ifthenelse{\equal{#2}{}}{}{,#2})}}
\newcommand{\CnsTempFuncSecond}[2] {\ensuremath{\CnsTempSecond(#1\ifthenelse{\equal{#2}{}}{}{,#2})}}

\newglossaryentry{cnstype}{%
	name={type},description={the type of a \glsentrydesc{cnstemp}}}
\def\CnsTempRep {\ensuremath{\mathfrak{C}}}
\newglossaryentry{cnsrep}{name={repertoire},description={the repertoire of \glsentrytext{declare} \glsentrytext{temp}s},
	symbol={\CnsTempRep}}
\newglossaryentry{cnsuniv}{name={\glsentrytext{con}s universe},description={the set of \glsentrytext{declare} \glsentrytext{temp}s over the process alphabet reflected in the log}}
\def\CnsInstRelation {\ensuremath{\Gamma}}

\newglossaryentry{cnsinst}{%
	name={\glsentrytext{cnstemp} instantiation relation},description={the assignment relation instantiating \glsentrytext{cnstemp}s into \glsentrytext{con}s, namely assigning \glsentrytext{task}s to \glsentrytext{cnspar}s.},
	symbol={\CnsInstRelation}}

\def\RelaConTemp {\ensuremath{\mathcal{R}}}
\newglossaryentry{relacontemp}{%
	name={relation template},description={the template of a relation \glsentrydesc{con}},
	symbol={\RelaConTemp}}

\def\ExiConTemp {\ensuremath{\mathcal{E}}}
\newglossaryentry{exicontemp}{%
	name={existence template},description={the template of an existence \glsentrydesc{con}},
		symbol={\ExiConTemp}}

\def\Supp {\ensuremath{\sigma}}

\newglossaryentry{support}{%
	name={support},description={the support of a \glsentrydesc{con}},
	symbol={\Supp}}

\def\Conf {\ensuremath{\kappa}}
\newglossaryentry{conf}{%
	name={confidence},description={the confidence level of a \glsentrydesc{con}},
	symbol={\Conf}}

\def\IntF {\ensuremath{\iota}}
\newglossaryentry{intf}{%
	name={interest factor},description={the interest factor of a \glsentrydesc{con}},
	symbol={\IntF}}

\def\EvaluationFunctor {\ensuremath{\eta}}
\newglossaryentry{evaluation}{
	name={evaluation},description={evaluation of a \glsentrytext{con} or a \glsentrytext{declamodel} over a \glsentrytext{evttrace} or an \glsentrytext{evtlog}},
	symbol={\EvaluationFunctor}}
\newcommand{\EvaluationOfOver}[2] {\ensuremath{\EvaluationFunctor\left(#1,#2 \,\right)}}

\def\PartTxt {Participation}
\def\UniqTxt {AtMostOne}

\def\ResExTxt {RespondedExistence}
\def\RespTxt {Response}
\def\RespTxtShort {Resp.}
\def\AltRespTxt {AlternateResponse}
\def\AltRespTxtShort {Alt.Response}
\def\ChaRespTxt {ChainResponse}
\def\PrecTxt {Precedence}
\def\PrecTxtShort {Prec.}
\def\AltPrecTxt {AlternatePrecedence}
\def\AltPrecTxtShort {Alt.Precedence}
\def\ChaPrecTxt {ChainPrecedence}
\def\CoExiTxt {CoExistence}
\def\SuccTxt {Succession}
\def\AltSuccTxt {AlternateSuccession}
\def\AltSuccTxtShort {Alt.Succession}
\def\ChaSuccTxt {ChainSuccession}

\def\ResExTmp {\ensuremath{\textsc{\ResExTxt}}}
\def\RespTmp {\ensuremath{\textsc{\RespTxt}}}

\def\PrecTmp {\ensuremath{\textsc{\PrecTxt}}}

\def\ChaSuccTmp {\ensuremath{\textsc{\ChaSuccTxt}}}

\newcommand{\Part}[1] {\ensuremath{\textsc{\PartTxt}(#1)}}
\newcommand{\Uniq}[1] {\ensuremath{\textsc{\UniqTxt}(#1)}}

\newcommand{\ResEx}[2] {\ensuremath{\textsc{\ResExTxt}(#1,#2)}}
\newcommand{\Resp}[2] {\ensuremath{\textsc{\RespTxt}(#1,#2)}}
\newcommand{\RespShort}[2] {\ensuremath{\textsc{\RespTxtShort}(#1,#2)}}

\newcommand{\AltResp}[2] {\ensuremath{\textsc{\AltRespTxt}(#1,#2)}}
\newcommand{\AltRespShort}[2] {\ensuremath{\textsc{\AltRespTxtShort}(#1,#2)}}
\newcommand{\ChaResp}[2] {\ensuremath{\textsc{\ChaRespTxt}(#1,#2)}}

\newcommand{\Prec}[2] {\ensuremath{{\textsc{\PrecTxt}}(#1,#2)}}
\newcommand{\PrecShort}[2] {\ensuremath{{\textsc{\PrecTxtShort}}(#1,#2)}}
\newcommand{\AltPrec}[2] {\ensuremath{\textsc{\AltPrecTxt}(#1,#2)}}
\newcommand{\AltPrecShort}[2] {\ensuremath{\textsc{\AltPrecTxtShort}(#1,#2)}}
\newcommand{\ChaPrec}[2] {\ensuremath{\textsc{\ChaPrecTxt}(#1,#2)}}
\newcommand{\CoExi}[2] {\ensuremath{\textsc{\CoExiTxt}(#1,#2)}}
\newcommand{\Succ}[2] {\ensuremath{\textsc{\SuccTxt}(#1,#2)}}
\newcommand{\AltSucc}[2] {\ensuremath{\textsc{\AltSuccTxt}(#1,#2)}}
\newcommand{\AltSuccShort}[2] {\ensuremath{\textsc{\AltSuccTxtShort}(#1,#2)}}
\newcommand{\ChaSucc}[2] {\ensuremath{\textsc{\ChaSuccTxt}(#1,#2)}}

\def\MultiSetFunctor {\ensuremath{\mathbb{M}}}
\newglossaryentry{multiset}{
	name={multi-set},description={a collection possibly containing multiple units of the same element},
	symbol={\MultiSetFunctor}}

\def\PowerSetFunctor {\ensuremath{\mathbb{P}}}
\newglossaryentry{powerset}{
	name={power-set},description={the collection of sets generated by all combinations without repetition of elements in a set},
	symbol={\PowerSetFunctor}}

\def\AutomInitState {\ensuremath{s_0}}

\newglossaryentry{fsainit}{name={initial state},description={initial state of the automaton},
	symbol=\AutomInitState}

\newacronym{rnf}{RNF}{Reactive Normal Form}
\newacronym{rf}{RCon}{Reactive Constraint}
\newacronym{sepautset}{sep.aut.set}{separated automata set}
\def\TruthDegree {\ensuremath{\zeta}}

\newglossaryentry{truthdegree}{%
	name={interestingness degree},description={the degree of interestingness of a \glsentrydesc{evttrace} to a \glsentrydesc{con}.},
	symbol={\TruthDegree}}

\newacronym{ltlp}{LTLp}{Linear-time Temporal Logic with Past}
\def\ltlpf {\ensuremath{\textrm{LTLp}_f}}
\newacronym{ltlpf}{\ltlpf}{Linear-time Temporal Logic with Past on Finite Traces}
\newacronym{ldl}{LDL}{Linear Dynamic Logic}
\newacronym{ldlf}{LDL$_f$}{Linear Dynamic Logic over Finite Traces}
\def\separationDegree {\textit{D}}
\newglossaryentry{sepdegree}{%
	name={separation degree},description={number of triples of a \glsentrytext{sepautset}},
	symbol={\separationDegree}}

\newglossaryentry{sat}{name={satisfaction},description={verification of a formula on a structure}}
\newglossaryentry{fulfilment}{name={fulfilment},description={satisfaction of a constraint on a trace in which the activation occurs}}
\newglossaryentry{activator}{name={activator},description={the event that signals the occurrence of the activation in the trace}}

\newglossaryentry{mirrorimage}{name={mirror image},description={temporal formula obtained by replacing all its operators with their mirror images}}

\newacronym{drva}{DRVA}{Declarative Rules Variant Analysis}
\newacronym{declens}{DecLens}{Declarative Lens}
\newacronym{ourTechnique}{\glsentrytext{drva}}{\glsentrydesc{drva}}

\def\varAlog{\ensuremath{L_A}}
\def\varAmodel{\ensuremath{M_A}}
\def\varAevents{\ensuremath{e_A}}
\def\varAeval{\ensuremath{E_A}}
\def\varApermLog{\ensuremath{L^\textrm{m}_{A_i}}}
\def\varAencodedLog{\ensuremath{L^\textrm{m}_{A}}}

\def\varEncodedLog{\ensuremath{L^\textrm{m}}}

\def\varBlog{\ensuremath{L_B}}
\def\varBmodel{\ensuremath{M_B}}
\def\varBevents{\ensuremath{e_B}}
\def\varBeval{\ensuremath{E_B}}
\def\varBpermLog{\ensuremath{L^\textrm{m}_{B_i}}}
\def\varBencodedLog{\ensuremath{L^\textrm{m}_{B}}}

\def\varUnionModel{\ensuremath{M_\cup}}
\def\varUnionDiff{\ensuremath{E_{\text{diff}}}}

\newcommand{\varGetFunction}[1]{\ensuremath{#1}}
\newcommand{\varGet}[2] {\ensuremath{\varGetFunction{#1}(#2)}}

\def\varRule{\ensuremath{r}}
\def\measureTxt{\ensuremath{m}}
\newcommand{\measure}[2] {\ensuremath{\textsc{\measureTxt}(#1,#2)}}

\def\varMindiff{\ensuremath{m_\text{diff-min}}}
\def\varMinMeasure{\ensuremath{m_\text{min}}}

\def\varPermutationsNum{\ensuremath{\pi}}
\def\varPermutationI{\ensuremath{i}}

\def\varCounter{\ensuremath{C}}

\def\pvalue{\ensuremath{\text{\textit{p}-value}}}
\def\significantLevel{\ensuremath{\alpha}}

\def\varHierarchyLevelMax{\ensuremath{h}}

\def\varDisplayedRules{\ensuremath{N}}

\def\varRuleExampleActivator{\taskize{ER\ Triage}}
\def\varRuleExampleTarget{\taskize{LacticAcid}}

\def\varRuleExampleShort{\Resp{\varRuleExampleActivator}{\varRuleExampleTarget}}
\usepackage{times}

\graphicspath{{images/}}

\begin{document}

\title{Detection of statistically significant differences between process variants through declarative rules}
\author{	Alessio Cecconi\inst{1} \and 
			Adriano Augusto\inst{2} \and 
			Claudio Di Ciccio\inst{3} 
}
\titlerunning{Process variants through declarative rules}

\institute{
           Vienna University of Economics and Business, Austria\\
             \hyperurl{mailto:cecconi@ai.wu.ac.at}{cecconi@ai.wu.ac.at}
           \and
           The University of Melbourne, Australia\\
             \hyperurl{mailto:a.augusto@unimelb.edu.au}{a.augusto@unimelb.edu.au}
           \and
           Sapienza University of Rome, Italy\\
             \hyperurl{mailto:claudio.diciccio@uniroma1.it}{claudio.diciccio@uniroma1.it}
}

\maketitle

\begin{abstract}
Services and products are often offered via the execution of processes that vary according to the context, requirements, or customisation needs. The analysis of such process variants can highlight differences in the service outcome or quality, leading to process adjustments and improvement. 
Research in the area of process mining 
has provided several methods for process variants analysis. However, 
very few of those account for a statistical significance analysis of their output. 
Moreover, those techniques 
detect differences at the level of process traces, single activities, or performance. 
In this paper, we aim at describing the distinctive behavioural characteristics between variants expressed in the form of declarative process rules.  
The contribution to the research area is two-pronged: the use of declarative rules for the explanation of the process variants and the statistical significance analysis of the outcome.
We assess the proposed method 
by comparing its results to the most recent process variants analysis methods. Our results demonstrate not only that declarative rules 
reveal differences at an unprecedented level of expressiveness, but also that our method outperforms the state of the art in terms of execution time.

\end{abstract}

\section{Introduction}\label{sec:intro}
%
The execution of a business process varies according to the context in which it operates. The exhibited behaviour changes to adapt to diverse requirements, geographical locations, availability or preferences of the actors involved, and other environmental factors. The alternative enactments lead to the diversification of specialised processes that stem from a general model. Considering some real-world examples, the hospital treatment of sepsis cases follows a different clinical pathway according to their age; the credit collection of road traffic fines typically undergoes an additional appeal to the prefecture when the fine is high.

Recent trends in business process management have led to a proliferation of studies that tackle the automated analysis of process variants~\cite{taymouri2021business}. After the initial manual, ad-hoc endeavours on single case studies, the focus has shifted towards the data-driven detection of their behavioural differences from the respective event logs~\cite{van2016log,bolt2018process,nguyen2018multi,taymouri2020business}.
%
To date, however, the explanatory power of the existing techniques is limited: the results are
\begin{iiilist}
	\item exposed as whole graphical models (e.g., directly-follows graphs) leaving the identification of the differences to the visual inspection of the analyst, or
	\item expressed as variations within the limited scope of subsequent event pairs (e.g., behavioural profiles).
\end{iiilist}

To overcome this limitation, we propose an approach that infers and describe the  
differences between variants in terms of behavioural rules. Our declarative approach aims to
\begin{iiilist}
	\item single out the distinct behavioural characteristics leading to the variations observed in the trace, while
	\item having a global scope as rules are exerted on the whole process runs.
\end{iiilist}
In our pursuit of explainability, we employ and adapt state-of-the-art techniques to guarantee the statistical significance of the inferred differentiating rules and expose the top-ranked distinctive characteristics in the form of natural language. We name our approach \gls{ourTechnique}.

In the following, \cref{sec:background} illustrates existing work in the areas upon which our approach is based. \Cref{sec:approach} describes in details our technique. \Cref{sec:eval} illustrates the results of our tool running on real-world event logs and compares them with state-of-the-art techniques. Finally, \cref{sec:conclusion} concludes this paper and provides remarks for future research avenues.
\section{Background and Related Work}\label{sec:background}

In this section, we provide a brief overview of the use of declarative languages within the process mining area. We discuss the most recent studies on process variants analysis, and we introduce the statistical significance test that we adapted to our context. 

\subsection{Declarative process specification and mining}
\begin{table}[tb]
	\caption{Description and notation of considered {\Declare} constraints}
	\label{tab:declare:explained}
	\centering
	\begin{tiny}
\begin{tabular}{ l l c c}
	\toprule
	\textbf{Constraint} & 
	\textbf{Verbal explanation}
	\\
	\midrule
	
	$\Part{\taska} $ & 
	$\taska$ occurs at least \emph{once}
	\\
	
	$\Uniq{\taska}$ & 
	$\taska$ occurs at most \emph{once}
	\\
	
	\midrule
	$\ResEx{\taska}{\taskb}$ &
	If {\taska} occurs, then {\taskb} occurs as well
	\\
	$\Resp{\taska}{\taskb}$ &
	If {\taska} occurs, then {\taskb} occurs after {\taska}
	\\
	$\AltResp{\taska}{\taskb}$ &
	Each time {\taska} occurs, then {\taskb} occurs afterwards, and no other {\taska} recurs in between
	\\
	$\ChaResp{\taska}{\taskb}$ &
	Each time {\taska} occurs, then {\taskb} occurs immediately afterwards
	\\
	$\Prec{\taska}{\taskb}$ &
	{\taskb} occurs only if preceded by {\taska}
	\\
	$\AltPrec{\taska}{\taskb}$ &
	Each time {\taskb} occurs, it is preceded by {\taska} and no other {\taskb} recurs in between
	\\
	$\ChaPrec{\taska}{\taskb}$ &
	Each time {\taskb} occurs, then {\taska} occurs immediately beforehand
	\\
	\midrule
	$\CoExi{\taska}{\taskb}$ &
	If {\taskb} occurs, then {\taska} occurs, and vice versa
	\\
	$\Succ{\taska}{\taskb}$ &
	{\taska} occurs if and only if it is followed by {\taskb}
	\\
	$\AltSucc{\taska}{\taskb}$ &
	{\taska} and {\taskb} if and only if the latter follows the former, and they alternate each other
	\\
	$\ChaSucc{\taska}{\taskb}$ &
	{\taska} and {\taskb} occur if and only if the latter immediately follows the former
	\\
	\bottomrule
\end{tabular}
	\end{tiny}
\end{table}
A declarative specification represents the behaviour of a process by means of reactive constraints~\cite{Cecconi2020ATemporal}, i.e., temporal rules that specify the conditions under which activities can or cannot be executed. 
For the purposes of this paper, we focus on {\Declare}, one of the most well-established declarative process modelling languages to date~\cite{pesic2010enacting}.
{\Declare} provides a standard repertoire of templates, namely linear temporal-logic rules parameterised over tasks, here abstracted as symbols of a finite non-empty alphabet $\LogAlph$.
\Cref{tab:declare:explained} illustrates the templates we use in the context of this paper.
Declarative rules are typically expressed in an if-then fashion, whereby the ``if'' part is named \emph{activator} and the ``then'' part is the \emph{target}.
For example, {\Resp{\taska}{\taskb}} requires that if {\taska} is executed (activator), then {\taskb} must be eventually carried out (the target).
{\Prec{\taskb}{\taska}} imposes that {\taska} cannot be executed if {\taskb} has not occurred earlier in the process instance.
{\ResEx{\taska}{\taskb}} relaxes the condition exerted by {\Resp{\taska}{\taskb}} and {\Prec{\taskb}{\taska}} by requiring that if {\taska} is executed, then {\taskb} has to occur in the same trace, regardless of whether it happens before or after {\taska}. {\Resp{\taska}{\taskb}} and {\Prec{\taskb}{\taska}} thus \emph{entail} {\ResEx{\taska}{\taskb}} and we say that {\ResExTmp} \emph{subsumes} {\RespTmp} and {\PrecTmp}~\cite{DBLP:journals/tmis/CiccioM15}.
Subsumption is a partial order on the templates of {\Declare} inducing the partial order of entailment on the rules that instantiate those templates.
A \emph{declarative specification} $M \ni r$ is a set of rules $r$ that conjunctively define the behaviour of a process.
We shall denote the universe of declarative rules as $\mathfrak{R} \supseteq M \ni r$.

The formal semantics of {\Declare} rules are rooted in \gls{ltlf}. 
A trace $t \in \LogAlph^*$ is a finite string of \emph{events}.
An event log (or \emph{log} for short) is a multi-set of traces $\EvtLog : \LogAlph^* \to \mathbb{N}$ (in which equivalent elements can occur multiple times).
We indicate the universe of event logs as $\mathfrak{L} \ni \EvtLog$.
We shall denote as $|t|$ the length of a trace $t$, as $|\EvtLog|$ the cardinality of the event log, and as $|\hat{t}|$ the length of the longest trace in the log.
Declarative process discovery tools can assess to what degree constraints hold true in a given event log.
To that end, diverse measures $\measureTxt : \mathfrak{R} \times \mathfrak{L} \to \mathbb{R}$ have been introduced to map a rule $r \in \mathfrak{R}¸$ and an event log $L \in \mathfrak{L}$ to a real number~\cite{Cecconi2020ATemporal}.
Among them, we consider \emph{Support} and \emph{Confidence} here.
Their values range from \num{0} to \num{1}.
Support is computed as the fraction of events in which both activator (e.g., the occurrence of {\taska} for {\Resp{\taska}{\taskb}}) and target (e.g., the occurrence of {\taskb} eventually afterwards in the trace) hold true.
Confidence is the fraction of the events in which the rule holds true over the events in which the activator is satisfied.
With a slight abuse of notation, we shall indicate the measure of $r$ on a single trace $t$ with $\measure{r}{t}$.

\subsection{Process variants analysis}

The latest literature review on process variants analysis~\cite{taymouri2021business} identified more than thirty studies addressing the research problem, clearly showing a growing interest in the topic. Much of the early research work was centred on case studies, providing examples of variant analysis applications that highlighted its potential capabilities. Among the earliest studies, Poelmans et al.~\cite{poelmans2010combining} combined process mining and data mining techniques to differentiate and analyse the healthcare pathways of patients affected by breast cancer and their response to therapies. Another of the most impactful studies 
is the one regarding a large Australian insurance company~\cite{suriadi2013understanding}, in which Suriadi et al.\ describe a methodology to pinpoint the weaknesses in the processing of insurance claims that had slow turnaround time.
Until recently, the vast majority of variant analysis studies focused on the detection of control flow differences of two or more process variants, with the frequent underlying goal to relate such differences to the outcome of the process variant. Also, proposed methods for variant analysis did not put much attention to the statistical significance of the detected differences, with the risk to catch random differences. Only recently proposed methods for variant analysis~\cite{bolt2018process,nguyen2018multi,taymouri2020business} account for statistical significance.

Bolt et al.~\cite{bolt2018process} pioneered the introduction of statistical significance in variant analysis, designing a framework to compare process variants in the form of annotated transition systems through the application of statistical tests. Also, the framework allows the analysis of the decision points in different variants and how the underlying decision-making rules differ. The annotations on the transition systems can capture frequencies and elapsed time between the transitions, however, the framework was designed to be extended, with the option to capture new annotations.

Nguyen et al.~\cite{nguyen2018multi} propose a variant analysis method based on the statistical comparison of \emph{perspective graphs}. A perspective graph is an artifact that captures an attribute of an event log (e.g. activity, resource, etc.), and the relations between the observed attribute values. The two perspective graphs (one per process variant) are then statistically compared and a \emph{differential perspective graph} is generated. The latter is analysed to determine the differences between the variants. By varying the attribute to generate the perspective graphs, the method allows for the analysis of different process perspectives.

Lastly, Taymouri et al.~\cite{taymouri2020business} propose a variant analysis method to separate statistically-significant different traces from common traces between the two input process variant logs. Each trace is encoded as a vector, then an SVM is used to classify the traces (assigning them to one of the two logs). A set of trace features (i.e. directly-follows relations) that can statistically discriminate the traces between the two logs is identified, and the logs are filtered by retaining only the traces containing those features. Finally, the two filtered logs are mapped into directly-follows graphs, called mutual-fingerprints, which rely on color-coding to show differences.

Of all the past studies, none is capable of analysing process variants to detect differences in the form of declarative rules~\cite{taymouri2021business}, which is instead the focus of this paper. However, given that declarative rules can be difficult to interpret, and that variant analysis is of high interest for the practitioners audience, we follow the example of van Beest et al.~\cite{van2016log} and provide a natural language translation of the detected differences~\cite{VanDerAa2020Say}.

\subsection{Statistical significance}
\label{sec:bg:significance}

It is likely that the data at hand captures only a portion of the whole data population, and it can include outliers. Process execution data is no exception. In fact, an event log usually contains only a part of the allowed process behaviour, on one hand, and infrequent behaviour, on the other hand~\cite{ProcessMiningBook2}.
To draw reliable observations from data samples, statistics provides several methods to assess the likelihood that an observation holds for the whole data population.
These methods, known as \emph{statistical significance tests}, estimate the probability that what is observed happened by chance. Traditionally, if such probability is low (below \num{0.05}), the observation is assumed to hold for the whole data population.


Of the many available methods to assess the statistical significance of the discovered declarative rules (which are similar to association rules)~\cite{hamalainen2019tutorial}, we deem the \emph{permutation test}~\cite{pitman1937significance} as the one that best fits for our purpose and context. 
The permutation test (also known as randomization test) is a non-parametric test originally designed to estimate the probability that two numerical series belong to the same population or not. In other words, it provides an 
answer to the question: ``Are the two series statistically different?''
The permutation test is carried out as follows. First, the difference of the averages of the two series is calculated (the \emph{delta average}). Then, the elements of the two series are pooled and two new series are drawn at random to compute the delta average again. This step is repeated for all the possible permutations. 
The likelihood that the two original series are statistically different is assessed as one minus the percentage of delta average that is greater or equal to the original delta average. 
The original permutation test is computationally heavy because it requires a computation on all the possible permutations of the original series. However, successive studies have proposed approximated~\cite{edgington1969approximate} and efficient~\cite{wu2016computing} methods to reduce the computational effort and construct permutation tests for any settings~\cite{welch1990construction}, from pure maths~\cite{pitman1937significance} to medicine~\cite{nichols2002nonparametric}. In the following, we show how we adapt the permutation test to address the problem of variant analysis.
\section{Approach}\label{sec:approach}
%
%
\begin{figure}[tb]
	\centering
	\includegraphics[width=1.0\linewidth]{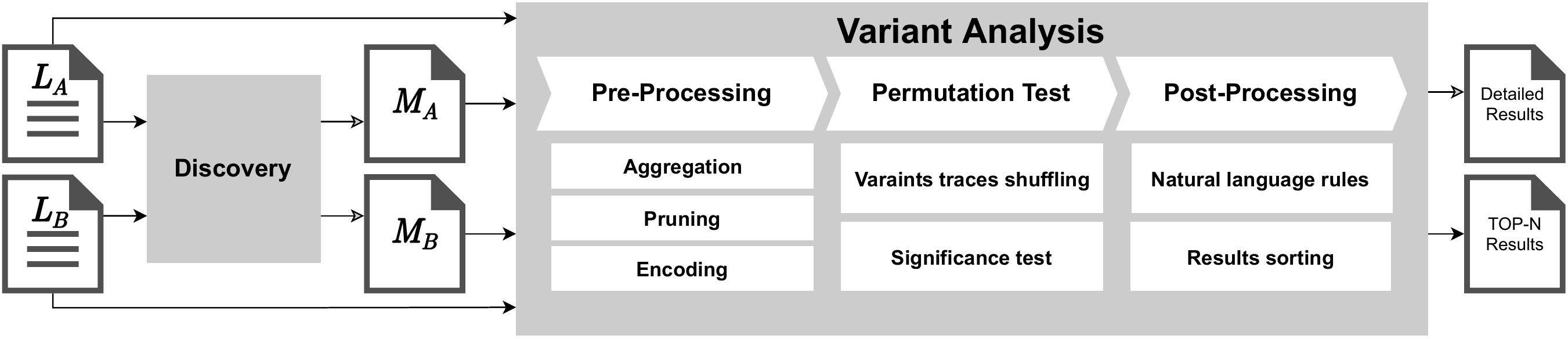}
	\caption[Method schema]{Schema of our variant analysis approach based on declarative rules. }
	\label{fig:method-schema}
\end{figure}
%
%
In this section, we present in details our variant analysis technique based on declarative process specifications: \acrfull{ourTechnique}.
The core idea is to discover the declarative process specifications describing the behaviour of the input variants' event logs and to check, by running a permutation test, whether the difference between the two specifications are statistically significant.
To assess the significance, intuitively, we verify whether the differences stemming from the variants' event logs are not due to rules that occur by chance. 

\Cref{fig:method-schema} presents the overview of the approach.
Specifically, the variant analysis is divided into four phases:
\begin{inparaenum}[(1)]
		\item A discovery phase, to mine the declarative specifications of the variants' behaviour;
        \item A pre-processing phase, to prune redundant or irrelevant rules and encode the information for faster computation; 
        \item A statistical validation phase, in which the permutations and related significance test take place; and 
        \item A final post-processing phase, to sort the results according to a given relevance criterion and produce a natural language description of the output.
\end{inparaenum}
In the remainder of this section, we thoroughly explain each step of the approach (illustrated as pseudo-code in \cref{algorithm:all}) and its implementation.

Some of the examples in this section refer to the publicly available SEPSIS event log (see \cref{tab:logs}), which records the treatment of patients with sepsis symptoms in a Dutch hospital. The two variants are generated as one containing traces regarding only elderly patients above the age of 70 (\varAlog) and the other containing young ones below 35 (\varBlog).
        
\begin{algorithm}[tb]
	\caption{Computing the set of rules that differ in a statistically significant way. The computational cost of the steps is specified on the right-hand side.}
	\label{algorithm:all}
	\begin{scriptsize}
   \SetAlgoLined
    \SetAlgoNoEnd
    \SetKwInOut{IN}{Input}
    \IN{Log variants {\varAlog} and {\varBlog}, specifications {\varAmodel} and {\varBmodel}, a function {\measureTxt}, parameters {\varMinMeasure}, {\varMindiff}, {\significantLevel}, {\varPermutationsNum}}
    \KwResult{$R$, a set of rules exhibiting a statistically significant difference between the variants.}
    \tcc{\textbf{Pre-processing}}
    $\varUnionModel \longleftarrow \varAmodel\cup\varBmodel$ \label{al:model-union} \tcp*{\tiny{Cost: $\mathcal{O}(|\varAmodel|+|\varBmodel|)$}}
    $\varUnionDiff,\varAeval, \varBeval \longleftarrow  \{\} $ \tcp*{\tiny{Cost: $\mathcal{O}(1)$}}
    \ForEach(\tcp*[f]{\tiny{Block cost: $\mathcal{O}(|\varUnionModel| \times |\hat{t}| \times ( |\varAlog| + |\varBlog|))$}}){$\varRule \in \varUnionModel$}{
        $ \varAeval \longleftarrow \varAeval \cup \{(r, \measure{\varRule}{\varAlog}) \}$ \label{al:pre-measuresA} \tcp*{\tiny{Cost: $\mathcal{O}(|\varAevents|+|\varAlog|)$}}
        $ \varBeval \longleftarrow \varBeval \cup \{(r, \measure{\varRule}{\varBlog}) \}$ \label{al:pre-measuresB} \tcp*{\tiny{Cost: $\mathcal{O}(|\varBevents|+|\varBlog|)$}}
        $ \varUnionDiff \longleftarrow \varUnionDiff \cup \{(r, |\measure{\varRule}{\varAlog}-\measure{\varRule}{\varBlog}|) \}$ \tcp*{\tiny{Cost: $\mathcal{O}(1)$}}
    }
    \ForEach(\tcp*[f]{\tiny{Block cost: $\mathcal{O}(|\varUnionModel|)$}}){$ \varRule\in\varUnionModel$}{
        \If(\label{al:pre-pruning}\tcp*[f]{\tiny{Cost: $\mathcal{O}(1)$}}){
            $(\varGet{\varUnionDiff}{\varRule} < \varMindiff) \lor 
            (\varGet{\varAeval}{\varRule}<\varMinMeasure) \lor
            (\varGet{\varBeval}{\varRule}<\varMinMeasure$)}{
            Remove {\varRule} from {\varUnionModel} and its measurements from {\varAeval}, {\varBeval}, {\varUnionDiff} \tcp*{\tiny{Cost: $\mathcal{O}(1)$}}
        }
    }
    $\varUnionModel,\varAeval,\varBeval,\varUnionDiff \longleftarrow$ hierarchicalSimplification(\varUnionModel,\varAeval,\varBeval) \label{al:simplification} \tcp*{\tiny{Cost: $\mathcal{O}(|\varUnionModel|\times|\varHierarchyLevelMax|)$}}
    $\varAencodedLog,\varBencodedLog \longleftarrow$ encodeLog(\varAlog,\varBlog,\varUnionModel) \tcp*{\tiny{Cost: $\mathcal{O}(|\varUnionModel|\times(|\varAlog|+|\varBlog|))$}}
    \tcc{\textbf{Permutation Test}}
    $ \varCounter \longleftarrow$ Initialize map such that for all $ \varRule\in\varUnionModel:  \varGet{\varCounter}{\varRule}=1$  \tcp*{\tiny{Cost: $\mathcal{O}(|\varUnionModel|)$}}
    \For(\label{al:permutations}\tcp*[f]{\tiny{Block cost: $\mathcal{O}(|\varPermutationsNum|\times|\varUnionModel|\times(|\varAlog|+|\varBlog|)) $}}){$i\leftarrow 1$ \KwTo \varPermutationsNum}{
        $\varApermLog,\varBpermLog \longleftarrow$ shuffleLog(\varAencodedLog,\varBencodedLog)  \label{al:shuffle} \tcp*{\tiny{Cost: $\mathcal{O}(|\varAlog|+|\varBlog|)$}}
        \ForEach(\tcp*[f]{\tiny{Block cost: $ \mathcal{O}(|\varUnionModel|\times(|\varAlog|+|\varBlog|)$}}){$ \varRule\in\varUnionModel$}{
            \If(\label{al:perm-measures}\tcp*[f]{\tiny{Cost: $\mathcal{O}(|\varAlog|+|\varBlog|)$}}){
            $|\measure{\varRule}{\varApermLog}-\measure{\varRule}{\varBpermLog}|\geq \varGet{\varUnionDiff}{\varRule}$}{
                $\varGet{\varCounter}{\varRule} \longleftarrow \varGet{\varCounter}{\varRule} + 1$ \tcp*[f]{\tiny{Cost: $\mathcal{O}(1)$}}
            }
        }
    }
    $ R \longleftarrow \{\}$ \tcp*{\tiny{Cost: $\mathcal{O}(1)$}}  
    \ForEach(\tcp*[f]{\tiny{Block cost: $\mathcal{O}(|\varUnionModel|)$}}){$ \varRule\in\varUnionModel$}{
        $\varGet{\pvalue}{\varRule} \longleftarrow \frac{\varGet{\varCounter}{\varRule}}{\varPermutationsNum}$ \tcp*{\tiny{Cost: $\mathcal{O}(1)$}}
        \lIf(\label{al:significance-check}\tcp*[f]{\tiny{Cost: $\mathcal{O}(1)$}}){$\varGet{\pvalue}{\varRule}\leq \significantLevel$}{
            $R \longleftarrow R \cup \{\varRule\}$
        }
    }
    \Return{$ R $}
\end{scriptsize}
\end{algorithm}

%
%
%
%

\subsection{Declarative specifications discovery}
\label{sec:discovery}
    In order to compare the variants through a declarative lens, it is necessary to derive the declarative specifications of their behavior. 
    Specifically, given the input variants event logs {\varAlog} and {\varBlog}, we want to retrieve the respective declarative specifications {\varAmodel} and \varBmodel.
    This can be done by executing existing declarative process discovery techniques~\cite{Slaats2020Declarative}.
    \todo{
    The discovery is not integrated into the statistical test, but it is in pipeline before it. 
    Thus our technique is independent from the specific declarative process discovery algorithm used for that purpose, because the only requirement is to produce a declarative process model for each input log. For example, in our experiments in~\cref{sec:eval}, we resort on MINERful~\cite{DBLP:journals/tmis/CiccioM15} as the fastest {\Declare} discovery algorithm~\cite{Slaats2020Declarative}, but other alternatives, e.g., \cite{Maggi2017,Cecconi2018Interestingness}, are seamlessly valid in our approach.
}
    
    We remark that the criteria used for the discovery influence the subsequent steps of the analysis.
    For example, discovering only specifications that are highly compliant to the event logs makes the variant analysis consider the most regular and stable behaviour of the processes. Looser specifications encompass also less frequent behaviours. 
    
    The discovery step is executed when only the event logs are given as input. Alternatively, our technique can receive as input the declarative specifications already discovered from the two logs, in such a case, the discovery step is skipped and our technique performs only the variant analysis with the input specifications. 
    Hand-crafted or filtered input declarative specifications can be useful when it is desired to test the statistical significance of the differences between two variants according to specific rules.
    For example, if the analyst is interested only in the behavioural difference involving tasks {\varRuleExampleActivator} and {\varRuleExampleTarget}, she can provide as input two specifications containing only rules involving these tasks (e.g., \Resp{\varRuleExampleActivator}{\varRuleExampleTarget}, \Prec{\varRuleExampleActivator}{\taskize{IV Liquid}}, \Succ{\varRuleExampleTarget}{\taskize{Admission NC}}, and so on) together with the variants' logs. 


\subsection{Pre-processing}
\label{sec:pre-processing}
    The variants analysis takes the variants' events logs {\varAlog} and {\varBlog} as input, together with the respective declarative process specifications {\varAmodel} and {\varBmodel}.
    As the permutation test is a computationally heavy task, it is desirable to keep only the essential information.
    Thus, to efficiently perform the statistical test, the data must be
    \begin{inparaenum}
        \item[\CircOne] aggregated, to have a common view between the variants, 
        \item[\CircTwo] cleaned via pruning, to remove redundant or irrelevant information, and 
        \item[\CircThree] encoded, to improve performance. 
    \end{inparaenum}
    
    \subsubsection{\CircOne~Aggregation.}
    \label{sec:aggregation}
    \begin{sloppypar}
    The declarative specifications {\varAmodel} and {\varBmodel} are merged into a unique specification $\varUnionModel=\varAmodel \cup \varBmodel$. To check their differences, all the rules in {\varAmodel} must be checked in {\varBlog} and vice versa, thus the union of {\varAmodel} and {\varBmodel} allows us to consider all and only the distinct rules in both the logs.\todo{clarify X2}

    \noindent
    The interestingness of each rule $\varRule\in\varUnionModel$ is measured in each variant log {\varAlog} and \varBlog.
    We resort to the measurement framework for declarative specifications proposed in~\cite{Cecconi2020ATemporal}. 
    Among the various measures available, we consider Confidence as the best option because it measures the degree of satisfaction of a rule in a log independently from the rule frequency.
    The comparison of the most suitable measures (or combinations thereof) goes beyond the scope of the paper and paves the path for future work.
    %
    In our example,
    $\measure{\varRuleExampleShort}{\varAlog}=0.83$
    and
    $\measure{\varRuleExampleShort}{\varBlog}=0.53$,
    which means that \SI{83}{\percent} of the occurrences of {\varRuleExampleActivator} in {\varAlog} are eventually followed by {\varRuleExampleTarget}. In {\varBlog}, the measure drops to \SI{53}{\percent} for that rule.
    \end{sloppypar}

    \noindent
    \begin{sloppypar}
    Finally, we retain these reference measurements 
    of {\varUnionModel} in each variant denoting them as functional relations {\varAeval} and \varBeval.
    Formally, we define 
    $ {\varAeval : \varUnionModel \to \mathbb{R}} $ as $ { \varAeval = \{ (\varRule, \measure{\varRule}{\varAlog}): \varRule \in \varUnionModel \} } $ 
    and
    $ {\varBeval : \varUnionModel \to \mathbb{R}} $ as $ { \varBeval = \{ (\varRule, \measure{\varRule}{\varBlog}): \varRule \in \varUnionModel \} } $. 
    We compute also the reference absolute difference between the variants' measurements {\varUnionDiff}.
    Formally, we define 
    $ {\varUnionDiff : \varUnionModel \to \mathbb{R}} $ as $ { \varUnionDiff = \{ (\varRule, |\measure{\varRule}{\varAlog} - \measure{\varRule}{\varBlog}|) : \varRule \in \varUnionModel \} } $, 
    where $|x|$ is the absolute value of $x$. 
    %
    For example,
    $\varGet{\varAeval}{\varRuleExampleShort}=0.83$,
    $\varGet{\varBeval}{\varRuleExampleShort}=0.53$, and their absolute difference is
    $\varGet{\varUnionDiff}{\varRuleExampleShort}=0.30$.
    \end{sloppypar}
    \subsubsection{\CircTwo~Pruning.}
    \label{sec:pruning}
    Not all the rules contained in {\varUnionModel} are valuable for the statistical test. Specifically, we consider as ignorable those rules that do not meet one of the following criteria. 
    
\begin{figure}[tb]
        \centering
        \resizebox{\textwidth}{!}{
            \begin{tikzpicture}[
        bad/.style={rectangle, draw=red!60, fill=red!5, very thick, minimum size=5mm,align=center},
        neutral/.style={rectangle, draw=gray!60, fill=gray!5, very thick, minimum size=5mm,align=center},
        good/.style={rectangle, draw=green!60, fill=green!5, very thick, minimum size=5mm,align=center},
        goodish/.style={rectangle, draw=orange!60, fill=orange!5, very thick, minimum size=5mm,align=center},
        node distance = 0.5cm
        ]
        
        \node[bad]   (AltSuccAB)                            {$\varRule_1 = \AltSucc{\taskt}{\taskv}$\\$\measure{\varRule_1}{\varAlog}=0.82 \qquad \measure{\varRule_1}{\varBlog}=0.49$};
        \node[neutral]   (AltRespAB)    [right=of AltSuccAB] {$\varRule_3 = \AltResp{\taskt}{\taskv}$\\$\measure{\varRule_3}{\varAlog}=0.82 \qquad \measure{\varRule_3}{\varBlog}=0.49$};
        \node[neutral]   (AltPrecAB)    [left=of AltSuccAB]  {$\varRule_2 = \AltPrec{\taskt}{\taskv}$\\$\measure{\varRule_2}{\varAlog}=1.00 \qquad \measure{\varRule_2}{\varBlog}=1.00$};
        
        \node[neutral]   (SuccAB)   [above=of AltSuccAB] {$\varRule_4 = \Succ{\taskt}{\taskv}$\\$\measure{\varRule_4}{\varAlog}=0.82 \qquad \measure{\varRule_4}{\varBlog}=0.49$};
        \node[neutral]   (RespAB)   [above=of AltRespAB] {$\varRule_6 = \Resp{\taskt}{\taskv}$\\$\measure{\varRule_6}{\varAlog}=0.82 \qquad \measure{\varRule_6}{\varBlog}=0.49$};
        \node[neutral]   (PrecAB)   [above=of AltPrecAB] {$\varRule_5 = \Prec{\taskt}{\taskv}$\\$\measure{\varRule_5}{\varAlog}=1.00 \qquad \measure{\varRule_5}{\varBlog}=1.00$};
        
        \node[neutral]   (CoExAB)   [above=of SuccAB] {$\varRule_7 = \CoExi{\taskt}{\taskv}$\\$\measure{\varRule_7}{\varAlog}=0.82 \qquad \measure{\varRule_7}{\varBlog}=0.49$};
        \node[good]   (ResExAB)    [above=of RespAB] {$\varRule_9 = \ResEx{\taskt}{\taskv}$\\$\measure{\varRule_9}{\varAlog}=0.82 \qquad \measure{\varRule_9}{\varBlog}=0.49$};
        \node[goodish]   (ResExBA)    [above=of PrecAB] {$\varRule_8 = \ResEx{\taskv}{\taskt}$\\$\measure{\varRule_8}{\varAlog}=1.00 \qquad \measure{\varRule_8}{\varBlog}=1.00$};
        
        
        \draw[->, line width=0.5mm] (CoExAB.east) -- (ResExAB.west);
        \draw[->, line width=0.5mm] (CoExAB.west) -- (ResExBA.east);
        
        \draw[->, line width=0.5mm] (SuccAB.east) -- (RespAB.west);
        \draw[->, line width=0.5mm] (SuccAB.west) -- (PrecAB.east);   
        
        \draw[->, line width=0.5mm] (AltSuccAB.east) -- (AltRespAB.west);
        \draw[->, line width=0.5mm] (AltSuccAB.west) -- (AltPrecAB.east);
        
        \draw[->, line width=0.5mm] (AltPrecAB.north) -- (PrecAB.south);
        \draw[->, line width=0.5mm] (PrecAB.north) -- (ResExBA.south);
        
        \draw[->, line width=0.5mm] (AltRespAB.north) -- (RespAB.south);
        \draw[->, line width=0.5mm] (RespAB) -- (ResExAB.south);
                
        \draw[->, line width=0.5mm] (AltSuccAB.north) -- (SuccAB.south);
        \draw[->, line width=0.5mm] (SuccAB) -- (CoExAB.south);
        
\end{tikzpicture}
        }
        \caption[Test]{Example of hierarchical simplification where the less restrictive rule is preferred over the more restrictive one, given equivalent measurements.}
        \label{fig:hierarchy-example}
    \end{figure}
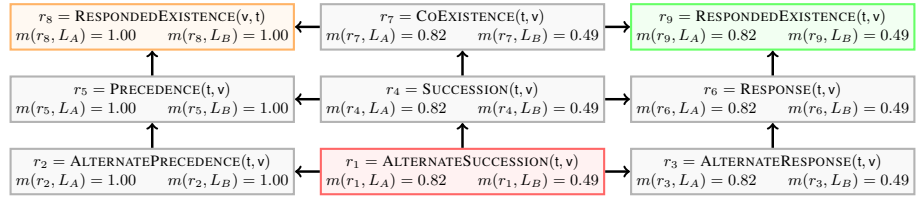

    \begin{asparadesc}
            \item[Minimum difference:] 
            If a rule difference between the variants is considered too small by an analyst to be of interest, it can be discarded. 
            %
            For example, rule
            $\varRule=\Prec{\taskize{ER\ Registration}}{\taskize{CRP}}$
            is such that 
            $\varGet{\varUnionDiff}{\varRule}=0.01$
            as
            $\varGet{\varAeval}{\varRule}=0.98$ and 
            $\varGet{\varBeval}{\varRule}=0.99$.
            The significance of such a difference is debatable: 
            %
            the difference appears to be negligible although the consideration is subjective and depending on the context of the analysis. 
            Therefore, we allow the user to customise a threshold to this end: {\varMindiff}. 
            \todo{Which step? (\cref{al:pre-pruning}) We may want to enrich the algorithm with comments emphasising the step.}
            According to this criterion, we remove
            all the rules $\varRule\in\varUnionModel$ such that $\varGet{\varUnionDiff}{\varRule} < \varMindiff$
            from {\varUnionModel} and all their measurements from {\varAeval}, {\varBeval}, and {\varUnionDiff}.

            \item[Minimum interestingness:] 
            If, according to an analyst, a rule is not interesting enough in either of the variants to  be  considered,  it  can  be  discarded.
            For example, rule
            $\varRule = \Resp{\taskize{ER\ Triage}}{\taskize{Release\ A}}$
            is such that
            $\varGet{\varUnionDiff}{\varRule}=0.23$ 
            yet the rule itself is not frequently satisfied in either of the variants as
            $\varGet{\varAeval}{\varRule}=0.41$ and
            $\varGet{\varBeval}{\varRule}=0.64$. 
            Whether this is desirable or not 
            depends on the context of the analysis.
            Therefore, we define the {\varMinMeasure} threshold to let the user set the desired minimum value that the rule's measure should be assigned within the variants' logs.
            In this step, we remove
            all the rules $\varRule' \in \varUnionModel $ such that
            $\measure{r'}{\varAlog} < \varMinMeasure$
            and
            $\measure{r'}{\varBlog} < \varMinMeasure$
            from {\varUnionModel} and all their measurements from {\varAeval}, {\varBeval}, and {\varUnionDiff}.
    
            \item[\textbf{No redundancy:}] If two rules are such that one is logically implied by the other, we do not gather additional information by retaining both in the set of rules under consideration. 
            {\Declare} patterns are hierarchically interdependent, as the satisfaction of a rule implies the satisfaction of all the entailed rules. 
            We can exploit the hierarchical relation of the {\Declare} templates to prune the redundant rules as in \cite{DiCiccio2017Resolving}.
            However, we adapt the original pruning technique by inverting the preference: We keep the most generic rule rather than the strictest one if measurements are the same in at least one of the variants' event logs. 
            \todo{Writing ``variants' event logs'' every time is tedious after a short while. I should write at the beginning that we simplify the phrasing speaking only of variants whenever clear from the context.}
            We explain the rationale at the core of our design choice with an example from the Sepsis event log.
            \begin{sloppypar}
            	\Cref{fig:hierarchy-example} depicts the partial order that stems from the subsumption relation among {\Declare} templates~\cite{DiCiccio2017Resolving} having
            	\AltSucc{\taskize{ER\ Sepsis\ Triage}}{\taskize{IV\ Antibiotics}}
            	as its least element,
            	and
            	\ResEx{\taskize{ER\ Sepsis\ Triage}}{\taskize{IV\ Antibiotics}}
            	and 
            	\ResEx{\taskize{IV\ Antibiotics}}{\taskize{ER\ Sepsis\ Triage}}
            	as its maximal elements.
            For the sake of space, we shall indicate \taskize{ER\ Sepsis\ Triage} and \taskize{IV\ Antibiotics} with {\taskt} and \taskv, respectively.
            \AltSuccShort{\taskt}{\taskv} is the strictest rule in the set, as all the other rules can be derived from it. As the associated Confidence is higher in {\varAlog} (\num{0.82}) than in {\varBlog} (\num{0.49}) we can claim the following: ``In {\varAlog}, {\taskv} follows {\taskt} and {\taskt} precedes {\taskv} with no other {\taskt} or {\taskv} occurring in between, more likely than in {\varBlog}''.
            Looking at the measurements on {\varAlog} and {\varBlog} we see that there is no difference between its values and the ones of \AltRespShort{\taskt}{\taskv} or \Succ{\taskt}{\taskv}, and it entails both.
            Recursively following the subsumption relation, we notice that the same measurements are associated to one of the maximal elements of the induced partial order: \ResEx{\taskt}{\taskv}.
            The other maximal element, instead, is such that the associated Confidence is equal to \num{1.00} in {\varAlog} and {\varBlog}.
            This characteristic reverberates along the chain of entailment down to \AltPrecShort{\taskt}{\taskv}.
            With this example, we observe that the least restrictive rules point out more precisely where the cause of the differences between variants lies -- in this case, the occurrence of {\taskv} required by {\taskt} is at the core of the distinct behaviours. The co-occurrence, order of execution, and lack of internal recurrence are characteristics that \AltPrecShort{\taskt}{\taskv} exhibits although they are evidenced by both variants.

            According to this criterion, we thus 
            prune redundant rules 
            only if the measurement between entailing and entailed rules are equivalent in at least one variant. Otherwise, we keep both. 
            Formally, denoting with $\vDash$ the entailment relation, we remove the following subset of rules from $\varUnionModel$ and all their measurements from {\varAeval}, {\varBeval}, and {\varUnionDiff}:
            $\{\varRule \in \varUnionModel : \varRule \vDash \varRule', \varRule' \in \varUnionModel \setminus \{ \varRule \} \textrm{ and } \measure{\varRule}{\varAlog}=\measure{\varRule'}{\varAlog} \textrm{ or } \measure{\varRule}{\varBlog}=\measure{\varRule'}{\varBlog} \}$.
            \end{sloppypar}
    \end{asparadesc}
    
	We remark that the first two criteria (minimum difference and minimum interestingness) are generically applicable to declarative rules and measures thereof, whereas the third one (no redundancy) is tailored for templates \'{a}-la-{\Declare} and measures such as Confidence, as it requires a subsumption hierarchy of the repertoire of templates and the monotone non-decrease of the measure within the subsumption hierarchy
	(i.e., if $\varRule \vDash \varRule'$, then $\measure{\varRule}{\EvtLog} \leqslant \measure{\varRule'}{\EvtLog}$) as shown in~\cite{DiCiccio2017Resolving}.


    \subsubsection{\CircThree~Encoding.}
    \label{sec:encoding}
    The measures of a trace in a log are based on the measurements of the rule in each trace of the log itself~\cite{Cecconi2020ATemporal}.
    Such trace measurement is independent from the log in which the trace resides, thus moving the trace from one log to another would not change it.
    We take advantage of this aspect to save computation steps: for each trace {\EvtTrace} in {\varAlog} and {\varBlog}, we cache the trace measurement \measure{\varRule}{\EvtTrace} of every rule $\varRule\in\varUnionModel$ in multi-sets {\varAencodedLog} and {\varBencodedLog}, i.e.,
    $\varAencodedLog=\{\measure{\varRule}{\EvtTrace}: \EvtTrace\in\varAlog, \varRule\in\varUnionModel \}$  and 
    $\varBencodedLog=\{ \measure{\varRule}{\EvtTrace}: \EvtTrace\in\varBlog, \varRule\in\varUnionModel \}$.
    \todo{ is it required to define also \EvaluationOfOver{\varRule}{\EvtTrace}, i.e., how a (reactive) rule is evaluated? Background?}
    This step allows us to encode the traces into feature vectors that can be used within the permutation test, as discussed in the next section.
    For example, given a trace {\EvtTrace} and a specification $\varUnionModel=\{\RespShort{\taska}{\taskb},\PrecShort{\taskc}{\taskd}\}$, the encoded trace is
    $ \EvtTrace^\textrm{m} = \{ (\RespShort{\taska}{\taskb}, \measure{\RespShort{\taska}{\taskb}}{\EvtTrace}), (\PrecShort{\taskc}{\taskd},\measure{\PrecShort{\taskc}{\taskd}}{\EvtTrace}) \}$.
    With a slight abuse of notation, we shall denote with $\measure{\varRule}{\varEncodedLog}$ the measure of a rule $\varRule$ on log $\EvtLog$ after the encoding of all traces $t \in \EvtLog$ in $\varEncodedLog$.
    %
    %
    
    \medskip\noindent
    To sum up, at the end of the pre-processing phase we have a unique declarative specification (\varUnionModel) 
    with the measurements in the variants (\varAeval, \varBeval, and \varUnionDiff),
    and the evaluation of each rule in every trace cached for later reuse ({\varAencodedLog} and \varBencodedLog).
    
\subsection{Permutation test}
\label{sec:permutation}
In this phase, we check whether the differences between the declarative rule measurements in the variants are statistically significant. 
In other words, taken each $\varRule\in\varUnionModel$, we calculate the likelihood that the absolute difference of its measurements between the variants, \varGet{\varUnionDiff}{\varRule}, was due to a random factor.
If the null hypothesis ``\varGet{\varUnionDiff}{\varRule} occurred by chance'' can be refuted, then the difference is significant.
To that extent, we employ an adaptation of the permutation test we introduced in \cref{sec:bg:significance}.
%
We reshuffle traces between {\varAlog} and {\varBlog} and observe if the difference in the measures holds as in the original variants' logs.
If so, it is likely that its difference was due to chance (i.e., the null hypothesis is confirmed). 
The rationale of this test is the following: if a difference can be detected by randomly shuffling the variants' traces, this difference has no real discriminative power between the variants. 

The acceptance or refutation of the null hypothesis is done in the following two steps (we continue the numbering from the previous section):
\begin{inparaenum}
 \item[\CircFour] The reshuffling, in which the data are rearranged and the measures of the rules under the null hypothesis are computed;
 \item[\CircFive] The significance test, in which only the rules that exhibit a statistically significant difference between variants are retained.
\end{inparaenum}
Notably, thanks to the encoding of the logs presented in~\cref{sec:encoding}, the trace evaluations of every rule are readily available to compute the measures across the new set of traces. Thus, we can shuffle the evaluations on the traces rather than the traces themselves.
Next, we elaborate on these operations.

    \subsubsection{\CircFour~Reshuffling.}
    \label{sec:shuffling}
        
	\begin{figure}[tb]
		\centering
		\begin{adjustbox}{width=1\textwidth,center=\textwidth}
			\begin{scriptsize}
				%
%
%
\begin{tabular}{l cc cc cc cc cc cc cc c cc}
	
	&
	\multicolumn{2}{c}{\includegraphics[width=1cm]{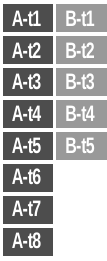}}
	&
	\multicolumn{2}{c}{\includegraphics[width=1cm]{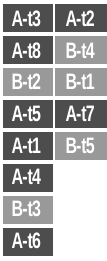}}
	&
	\multicolumn{2}{c}{\includegraphics[width=1cm]{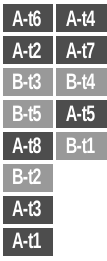}}
	&
	\multicolumn{2}{c}{\includegraphics[width=1cm]{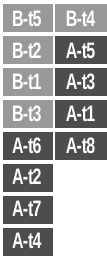}}
	&
	\multicolumn{2}{c}{\includegraphics[width=1cm]{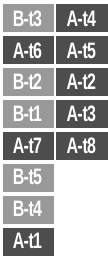}}
	&
	\multicolumn{2}{c}{\includegraphics[width=1cm]{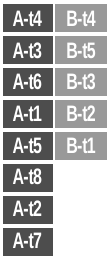}}
	&
	\multicolumn{2}{c}{\includegraphics[width=1cm]{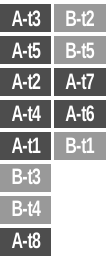}}
	&
	\multicolumn{1}{c}{$\dots$}
	&
	\multicolumn{2}{c}{\includegraphics[width=1cm]{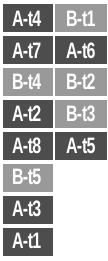}}
	\\
	
	\toprule
	&
	\multicolumn{1}{|c}{$ L_A $} &
	\multicolumn{1}{c}{$ L_B $} &
	\multicolumn{1}{|c}{$ L_{A'} $} &
	\multicolumn{1}{c}{$L_{B'}$} &
	\multicolumn{1}{|c}{$L_{A\dblprime}$} &
	\multicolumn{1}{c}{$L_{B\dblprime}$} &
	\multicolumn{1}{|c}{$L_{A\triprime}$} &
	\multicolumn{1}{c}{$L_{B\triprime}$} &
	\multicolumn{1}{|c}{$L_{A\qdrprime}$} &
	\multicolumn{1}{c}{$L_{B\qdrprime}$} &
	\multicolumn{1}{|c}{$L_{A^{(5)}}$} &
	\multicolumn{1}{c}{$L_{B^{(5)}}$} &
	\multicolumn{1}{|c}{$L_{A^{(6)}}$} &
	\multicolumn{1}{c|}{$L_{B^{(6)}}$} &
	\multicolumn{1}{c}{$\dots$} &
	\multicolumn{1}{|c}{$L_{A^{(1000)}}$} &
	\multicolumn{1}{c}{$L_{B^{(1000)}}$}
	\\
	
	\midrule	
	
	
	\measure{\varRule}{L} &
	 \multicolumn{1}{|c}{1.00} &
	 \multicolumn{1}{c}{0.00} &
	 \multicolumn{1}{|c}{0.75} &
	 \multicolumn{1}{c}{0.40} &
	 \multicolumn{1}{|c}{0.63} &
	 \multicolumn{1}{c}{0.60} &
	 \multicolumn{1}{|c}{0.50} &
	 \multicolumn{1}{c}{0.80} &
	 \multicolumn{1}{|c}{0.38} &
	 \multicolumn{1}{c}{1.00} &
	 \multicolumn{1}{|c}{1.00} &
	 \multicolumn{1}{c}{0.00} &
	 \multicolumn{1}{|c}{0.75} &
	 \multicolumn{1}{c|}{0.40} &
	 $\dots$ &
	 \multicolumn{1}{|c}{0.75} &
	 \multicolumn{1}{c}{0.40} 
	 \\
	
	\varUnionDiff & 
	\multicolumn{2}{|c}{1.00} & 
	\multicolumn{2}{|c}{0.35} &
	\multicolumn{2}{|c}{0.03} &
	\multicolumn{2}{|c}{0.30} &
	\multicolumn{2}{|c}{0.63} &
	\multicolumn{2}{|c}{1.00} &
	\multicolumn{2}{|c|}{0.35} &
	$\dots$ &
	\multicolumn{2}{|c}{0.35}\\

	\midrule
	&
	\multicolumn{5}{c}{$ \text{Permutations}=1000 $} &
	\multicolumn{3}{c}{$ \significantLevel= 0.01 $}	&	
	\multicolumn{4}{c}{$\varGet{\varCounter}{\varRule}=5$} &
	\multicolumn{5}{r}{$\varGet{\pvalue}{\varRule}=0.005 \implies \textbf{significant}$}
	\\
	
	\bottomrule
\end{tabular}
			\end{scriptsize}
		\end{adjustbox}
		\caption[Visual example of the permutation test.]{Visual example of the permutation test. The traces of the original logs {\varAlog} and {\varBlog} are coloured in dark gray and light grey, respectively.} 
		\label{fig:permutation-example}
	\end{figure}
    \Cref{fig:permutation-example}
    presents a graphical example of the reshuffling. At each iteration, the traces are randomly rearranged between the variants' logs, thereby generating altered logs on which the rule is measured.
    %
    Specifically, at each iteration {\varPermutationI}:
    \begin{iiilist}
        \item The multi-set of all encoded traces in ${\varAencodedLog} \cup {\varBencodedLog}$ is shuffled into two new logs {\varApermLog} and {\varBpermLog}, such that $|{\varAencodedLog}| = |{\varApermLog}|$ and $|{\varBencodedLog}| = |{\varBpermLog}|$ (permutation step);
        \item For each $\varRule\in\varUnionModel$ the measures \measure{\varRule}{\varApermLog} and \measure{\varRule}{\varBpermLog} are computed; 
        \item Finally, for each rule $\varRule\in\varUnionModel$ the difference of its measurements is compared to the reference difference of the rule \varGet{\varUnionDiff}{\varRule}.
    \end{iiilist}
    The number of iterations is set as a user parameter $\varPermutationsNum$: according to~\cite{edgington1969approximate}, a suitable value for $\varPermutationsNum$ is \num{1000}.
    We denote with $\varGetFunction{\varCounter}: \varUnionModel \to \mathbb{N}$ the function mapping a rule ${\varRule} \in \varUnionModel$ to the number of iterations in which its difference is greater than or equal to the reference difference, 
    i.e., $\varGet{\varCounter}{\varRule}=
    \sum_{\varPermutationI=1}^{\varPermutationsNum}{\llbracket |\measure{\varRule}{\varApermLog}-\measure{\varRule}{\varBpermLog}|\geq\varGet{\varUnionDiff}{\varRule} \rrbracket}$ where $\llbracket \cdot \rrbracket$ is the indicator function mapping to \num{1} or \num{0} if the argument holds true or false, respectively.

    \subsubsection{\CircFive~Significance test.}
    \label{sec:significance}
        At the end of the permutations step, the counter \varGet{\varCounter}{\varRule} tells us for each rule how frequently a difference greater or equal the reference one is observed.
%
        The likelihood to observe the rules difference under the assumption of the null hypothesis is the {\pvalue} of the test: $\varGet{\pvalue}{\varRule}=\frac{\varGet{\varCounter}{\varRule}}{\varPermutationsNum}$.
        The significance level {\significantLevel} of the test is the {\pvalue} threshold below which the null hypothesis should be discarded.
        It is common to set 
        ${\significantLevel} = 0.01$ when the permutation test consists of \num{1000} iterations~\cite{edgington1969approximate}.  
        %
        A rule $\varRule\in\varUnionModel$ has a statistically significant difference between the variants {\varAlog} and {\varBlog} if and only if $\varGet{\pvalue}{\varRule}\leq \significantLevel $.
        \Cref{fig:permutation-example} illustrates an example of such a significance test where a difference of at least 
        \varGet{\varUnionDiff}{\varRule} occurred for only 
        \num{5} permutations out of the \num{1000} performed, e.g, in permutation $ \varPermutationI = 5 $. 
        Therefore, 
        $\varGet{\pvalue}{\varRule}=0.005$. As it is less than the significance level of \num{0.01}, it suggest that the difference is statistically significant.

        \medskip
        \noindent
        In conclusion, the technique returns the set of rules that determine a statistically significant difference between the variants. 
        
    \paragraph{About the computational cost.}
    \label{sec:computational-cost}
        \Cref{algorithm:all} shows the pseudo-code of our approach for the pre-processing and the permutation test phases.
        The overall computational cost is linear in the input size. 
        Let 
        $|\varAmodel|$, $|\varBmodel|$, and $|\varUnionModel|$ be the number of rules in, respectively, \varAmodel, \varBmodel, and \varUnionModel, 
        $|\varAlog|$ and $|\varBlog|$ the number of traces in {\varAlog} and \varBlog, 
        $|\hat{t}|$ the maximum length of a trace in $\varAlog \cup \varBlog$,
        $\varHierarchyLevelMax$ the maximum hierarchical level of a rule in \varUnionModel, and 
        $\varPermutationsNum$ the number of iterations in the permutation test. 
        Summing the costs of each step of \cref{algorithm:all} (on the right side of every line),
        the overall computation cost of the approach is: 
            $ \mathcal{O}(|\varUnionModel| \times |\hat{t}| \times (|\varAlog| + |\varBlog|)) + \mathcal{O}(|\varUnionModel| \times \varPermutationsNum \times (|\varAlog| + |\varBlog|)) $.

        The overall performance is driven by the cost of the evaluation of the rules on the traces (the encoding phase) and by the permutation tests.
        We keep the two addenda separated to highlight their contribution.
        Depending on whether $|\hat{t}|$ is greater or less than $\varPermutationsNum$, the first or the second addendum prevails. 
        We remark that the cost of computing the measures in the pre-processing (\cref{al:pre-measuresA,al:pre-measuresB})
        stems from the computational cost of the measurement framework~\cite{Cecconi2020ATemporal}, 
        while the cost of re-computing the measures during the permutations (\cref{al:perm-measures}) subtracts the trace evaluation time to that cost due to the encoding presented in \cref{sec:encoding}.  
        For as far as the hierarchical simplification (\cref{al:simplification}), a {\Declare} rule from the standard templates set may have at most $h=11$  (the {\ChaSuccTmp} template). 

\subsection{Post-processing}
\label{sec:post-processing}
    Once the rules with a statistically significant difference between the variants have been identified, we show them to the final user. 
    All the relevant details are reported, namely the rule \varRule, its \pvalue, its original measurements in the variants logs \measure{\varRule}{\varAlog} and \measure{\varRule}{\varBlog}, and their absolute difference \varGet{\varUnionDiff}{\varRule}.
    Furthermore, in order to enhance the clarity of the outcome, we perform the following additional steps: 
    \begin{iiilist}
        \item providing a natural-language description of the output, and
        \item sorting the results according to a priority criterion.
    \end{iiilist}
    
    \subsubsection{\CircSix~Natural language.} We report the rules along with a natural language description explaining their behaviour in a concise manner.
    Indeed, the comprehension of temporal logic formulae is out of the reach for a general audience, and even the {\Declare} rules taken by their own are not immediately grasped by non-knowledgeable users~\cite{VanDerAa2020Say}. For example, \Resp{\varRuleExampleActivator}{\varRuleExampleTarget} is more readable than 
    its {\ltlpf} formulation, provided that the user has a prior knowledge of \Declare. 
    To explain differences between variants, we thus provide a description as follows: ``In variant $A$, it is 30\% more likely than in variant $B$ that if {\varRuleExampleActivator} occurs, {\varRuleExampleTarget} will occur afterwards''. \todo{check number with previous examples}
    
    \subsubsection{\CircSeven~Sorting.} We give higher priority to those rules whose measure differs the most between variants. 
    To ease the discoverability of prominent results, we sort the rules in descending order according to their measurement difference between the variants, \varGet{\varUnionDiff}{\varRule}, and, 
    in the case of equal difference, the highest measurement in {\varAlog} or {\varBlog}. 
    For the sake of readability, the user can specify how many of the ranked rules should be initially displayed by means of a dedicated parameter, $\varDisplayedRules$.
    

\subsection{Implementation and remarks}
\label{sec:implementation}
We have implemented our approach, \acrfull{ourTechnique}, as a Java command-line tool.%
\footnote{Available at: \url{https://github.com/Oneiroe/Janus}} 
Our tool returns two files: 
\begin{iiilist}
	\item a textual file containing the top-{\varDisplayedRules} statistically significant differences between the input logs reported in natural language, and
	\item one tabular file (CSV format) containing all the statistically-significant differences captured as declarative rules (and in natural language) with the entire quantitative information from the analysis. 
\end{iiilist}
As input, it takes two event logs ($\varAlog$ and $\varBlog$, in XES, MXML, or CSV formats), two declarative specifications ($\varAmodel$ and $\varBmodel$, in JSON or XML formats) and the following list of user-defined parameters to tune the algorithm: 
%
\begin{iiilist}
    \item {\measureTxt}, the measure to use for the comparison of the variants among those that are supported by the measurement framework described in~\cite{Cecconi2020ATemporal} (the default is Confidence, as explained in~\cref{sec:aggregation});
    \item {\varMinMeasure}, the minimum measure threshold that a rule should exceed in at least one variant (\num{0.00} by default); 
    %
    \item {\varMindiff}, the minimum difference threshold that a rule should exceed between the variants 
    (\num{0.01} by default); 
    %
    %
    \item {\varPermutationsNum}, the number of iterations to perform during the permutation test, 
    set to \num{1000} by default as explained in \cref{sec:shuffling}; 
    \item {\significantLevel}, the significance level, namely the maximum {\pvalue} to deem a rule as statistically relevant to discriminate the variant 
    (\num{0.01} by default); 
    \item {\varDisplayedRules}, the number of top rules to display in the 
    textual output (see \cref{sec:post-processing}), set to \num{10} by default.
\end{iiilist}
%

Our tool can be put in pipeline to a declarative discovery algorithm which mines the declarative specifications out of the variants' logs. 
In the experiment presented in following \cref{sec:eval}, we 
pass the output of the MINERful discovery algorithm~\cite{DBLP:journals/tmis/CiccioM15} 
to \gls{ourTechnique}.

\section{Evaluation}\label{sec:eval}

Given that no previous work addressed the problem of variant analysis via declarative rules~\cite{taymouri2021business}, in this section, we provide a qualitative comparison with the latest process variants analysis approach that discovers statistically-significant differences~\cite{taymouri2020business}, which already demonstrated its advantages~\cite{taymouri2020business} with respect to other baselines~\cite{bolt2018process,nguyen2018multi}. Henceforth, we will refer to 
the approach of Taymouri et al.~\cite{taymouri2020business} as MFVA (Mutual Fingerprints Variant Analysis).
Furthermore, we also consider the work of van Beest et al.~\cite{van2016log} as a baseline for comparison given that its output is in natural language statements -- henceforth referred to as PESVA (Prime Event Structure Variant Analysis). However, we note that PESVA does not take into account the statistical significance of the detected differences and it is not based on declarative rules either. Instead, PESVA outputs two type of statements:
\begin{iiilist}
	\item frequency-based statements, which highlight differences in branching probabilities of the process decision points; and
	\item behaviour-based statements, which highlight differences in directly-follow, and concurrent relations, as well as optional tasks.
\end{iiilist}

\subsubsection{Dataset and setup.}
\begin{table}[tb]
	\caption{Descriptive statistics of the evaluation dataset~\cite{taymouri2020business}}
	\label{tab:logs}
	\centering
	\resizebox{1.0\textwidth}{!}{%
		\begin{tabular}{l|l|rr|rr|rrr}

    \multicolumn{2}{c|}{\textbf{Log}}
    & \multicolumn{2}{c|}{\textbf{Traces}}
    & \multicolumn{2}{c|}{\textbf{Events}}
    & \multicolumn{3}{c}{\textbf{Trace length}}
    \\\hline
    \textbf{Name and DOI}
    & \textbf{Variant}
    & \textbf{Total}
    & \textbf{Distinct}
    & \textbf{Total}
    & \textbf{Distinct}
    & \textbf{Min}
    & \textbf{Avg}
    & \textbf{Max} \\\hline
	
	BPIC13 & Company = A\textsubscript{2} & 553 & 25.5\% & 4,221 & 3 & 2 & 8 & 53\\\cline{2-9}
    (\hyperurl{http://dx.doi.org/10.4121/uuid:a7ce5c55-03a7-4583-b855-98b86e1a2b07}{10.4121/uuid:a7ce5c55-03a7-4583-b855-98b86e1a2b07}) & Company = C & 4,417 & 13.8\% & 29,122 & 4 & 1 & 7 & 50\\\hline
	
	BPIC15 & Municipality = 1 & 1,199 & 97.6\% & 36,705 & 146 & 2 & 33 & 62\\\cline{2-9}
    (\hyperurl{http://dx.doi.org/10.4121/uuid:915d2bfb-7e84-49ad-a286-dc35f063a460}{10.4121/uuid:915d2bfb-7e84-49ad-a286-dc35f063a460}) & Municipality = 2 & 831 & 99.6\%  & 32,017 & 134 & 1 & 39 & 96\\\hline
    	   
	RTFMP & Fine Amount $\geq$ 50 & 21,243 & 0.7\%  & 91,499 & 11 & 2 & 4 & 20\\\cline{2-9}
    (\hyperurl{https://doi.org/10.4121/uuid:270fd440-1057-4fb9-89a9-b699b47990f5}{10.4121/uuid:270fd440-1057-4fb9-89a9-b699b47990f5}) & Fine Amount $<$ 50 & 129,127 & 0.1\% & 469,971 & 11 & 2 & 4 & 11\\\hline
    	   
	SEPSIS & Patient Age $\geq$ 70 & 678 & 85.7\%  & 10,243 & 16 & 3 & 15 & 185\\\cline{2-9}
    (\hyperurl{http://dx.doi.org/10.4121/uuid:915d2bfb-7e84-49ad-a286-dc35f063a460}{10.4121/uuid:915d2bfb-7e84-49ad-a286-dc35f063a460}) & Patient Age $\leq$ 35 & 76 & 67.1\%  & 701 & 12 & 3 & 9 & 52\\\hline
\end{tabular}

	}
\end{table}
We reproduce the experimental setup proposed in~\cite{taymouri2020business}.
The evaluation dataset consists of four publicly available event logs. Each of the logs can be divided into two variants (based on process-instance attribute values).
The descriptive statistics characterizing the four logs are shown in \cref{tab:logs}. 
The original logs (not partitioned by the variants) can be downloaded from the 4TU Research Data Centre.%
\footnote{\label{footnote:datasets}\url{https://data.4tu.nl/search?categories=13503}}
The process variant logs can be downloaded from our online repository 
together with the full set of results of this evaluation which, due to space limits, we could not report in this section.~\footnote{\url{https://github.com/Oneiroe/DeclarativeRulesVariantAnalysis-static}} In the following, we focus our discussion on the most interesting results.%

In our evaluation, we give each pair of log variants as an input to \gls{ourTechnique}, MFVA, and PESVA. While \gls{ourTechnique} and PESVA produce natural language statements, 
MFVA returns graphs named \emph{mutual fingerprints}.
Since the different outputs do not allow for a straightforward comparison, we first analyse the results of \gls{ourTechnique} and MFVA with the goal to highlight the commonalities and differences in their output, 
then we compare our output with that of PESVA. We remark that the three tools provide alternative perspectives of the process variants differences. Therefore, they should not be seen as mutually exclusive tools, but rather complementary. 

After the qualitative comparison, we also conduct a performance comparison of the three tools, reporting their execution times. All the experiments were run on an Intel Core i7-8565U@1.80GHz with 32GB RAM equipped with Windows 10 Pro (64-bit), with a timeout of 3 hours per variant analysis.
The input parameters we used for our \gls{ourTechnique} are the default ones explained in \cref{sec:implementation}. 
The declarative process models are discovered through MINERful~\cite{DBLP:journals/tmis/CiccioM15} with a Support threshold of \num{0.5} and a Confidence threshold of \num{0.0}, 
so as to keep the specifications loose enough to capture rare behaviour too.
Markedly, the execution times reported for \gls{ourTechnique} include the discovery times. 
%
%

\subsubsection{Results and discussion.}

\begin{figure}[tb]
	\begin{floatrow}
		\ffigbox{%
			\begin{subfigure}[t]{.24\textwidth}
				\centering
				\includegraphics[height=6cm]{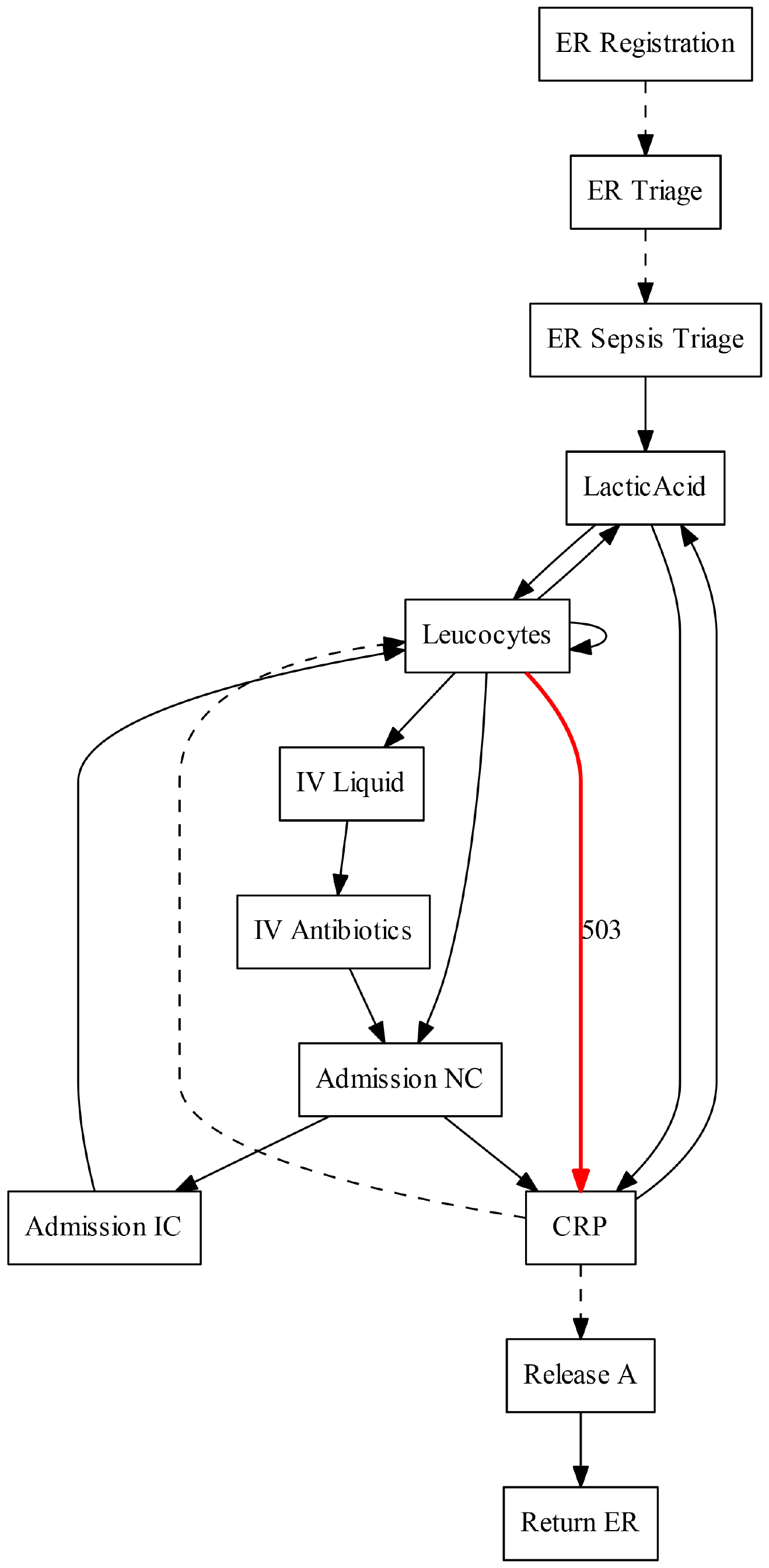}
				\caption{$\text{Age} \geq 70 $}
				\label{fig:sepsis:1}
			\end{subfigure}%
			\begin{subfigure}[t]{.24\textwidth}
				\centering
				\includegraphics[height=6cm]{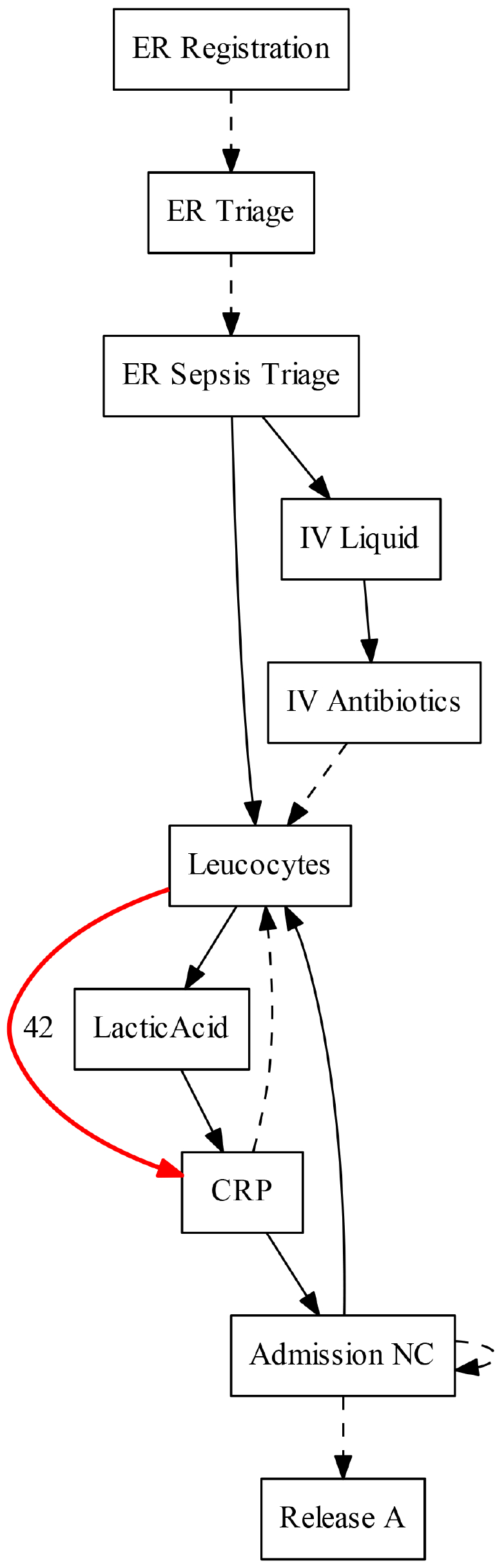}
				\caption{$\text{Age} \leq 35 $}
				\label{fig:sepsis:2}
			\end{subfigure}
		}{%
			\caption{SEPSIS logs, MFVA output~\cite{taymouri2020business}}
			\label{fig:sepsis}
		}
		\killfloatstyle
		\ffigbox{%
			\begin{subfigure}[t]{.25\textwidth}
				\centering
				\includegraphics[height=6cm]{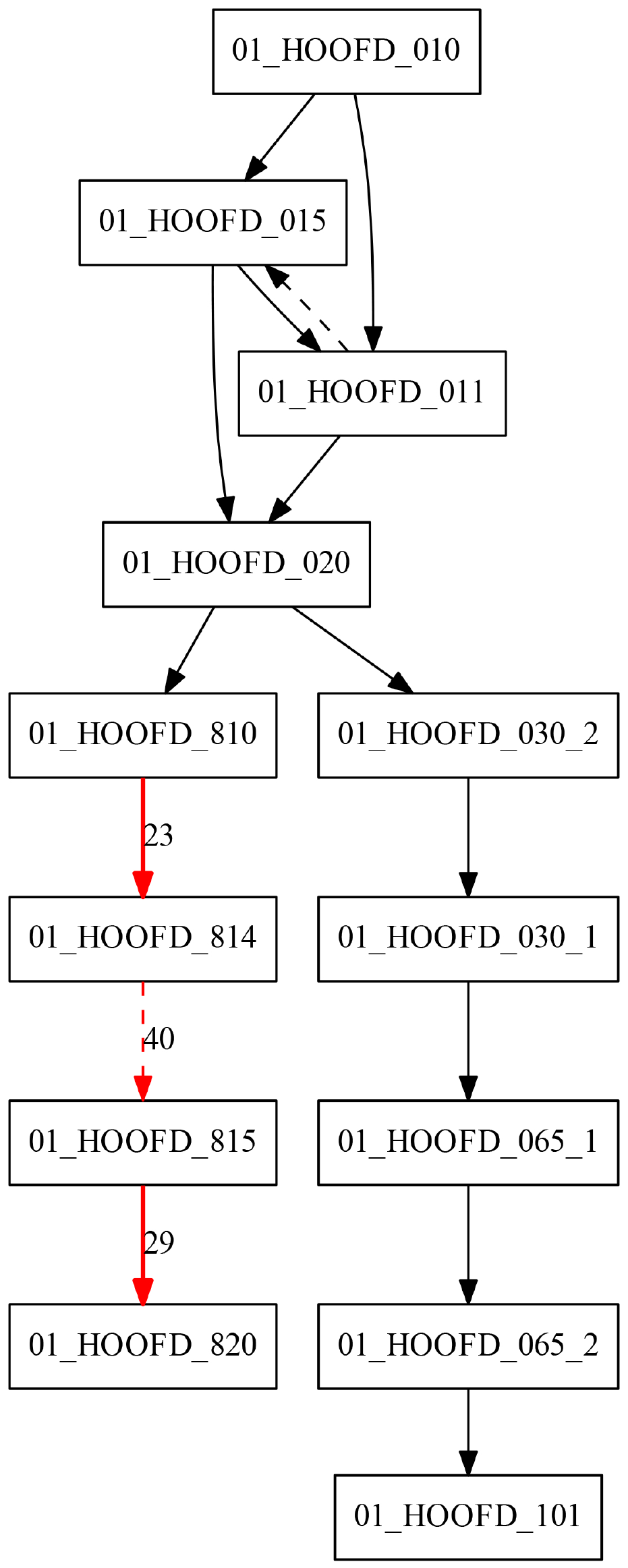}
				\caption{BPIC15 (M1)}
				\label{fig:bpic15:1}
			\end{subfigure}%
			\begin{subfigure}[t]{.25\textwidth}
				\centering
				\includegraphics[width=\textwidth, angle=90, height=6cm]{images/bpic15_2_TEMP-smaller.pdf}
				\caption{BPIC15 (M2)}
				\label{fig:bpic15:2}
			\end{subfigure}
		}{%
			\caption{BPIC15 logs, MFVA output~\cite{taymouri2020business}}
			\label{fig:bpic15} %
		}
	\end{floatrow}
\end{figure}
\Cref{fig:sepsis} shows the output of MFVA given the variants of the SEPSIS log as an input. In the figure, dashed arcs and red arcs capture a statistically significant difference. The former evidence a change in the processing-time between the traces of the two variants,
while the latter capture a 
difference in frequency (annotated on the arcs).
\Cref{tab:sepsis} shows the output of \gls{ourTechnique} given the same logs. 
We can notice that the difference ranked first in the table (\Part{\taskize{Admission NC}} is more likely to occur if the age of the treated patient is over \num{70}) makes explicit a difference that could not be directly inferred by visually comparing \cref{fig:sepsis:1} with \cref{fig:sepsis:2}, although the alternative paths may suggest a behavioural change. The same holds for the fifth and ninth evidence. We observe that the other statements are related to the three we mentioned (also highlighted in \cref{tab:sepsis}). A further simplification step to enhance the ranking according to the discriminative power of the statements is an intriguing problem that paves the path for future endeavours building upon this work.

\begin{table}[tb]
	\caption[Output from the Sepsis logs]{SEPSIS logs -- \gls{ourTechnique} output (our approach)}
	\label{tab:sepsis}
	\centering
	\resizebox{\textwidth}{!}{%
		\begin{tabular}{l|l}

    & \textbf{Statistically significant differences (\textit{Variant A}: $\mathrm{Age} \geq 70$; \textit{Variant B}: $\mathrm{Age} \leq 35$)} \\\hline
	\grayrow
	1 & In \textit{Variant A}, it is 37.4\% more likely than in \textit{Variant B} that \taskize{Admission NC} occurs in a process instance.\\\hline
    2 & In \textit{Variant A}, it is 37.4\% more likely than in \textit{Variant B} that if \taskize{ER Registration} occurs, \taskize{Admission NC} will occur\\ & afterwards without any other occurrence of \taskize{ER Registration} in between.\\\hline
    3 & In \textit{Variant A}, it is 37.4\% more likely than in \textit{Variant B} that if \taskize{ER Sepsis Triage} occurs, also \taskize{Admission NC} occurs.\\\hline
    4 & In \textit{Variant A}, it is 37.4\% more likely than in \textit{Variant B} that if \taskize{ER Triage} occurs, also \taskize{Admission NC} occurs.\\\hline
    \grayrow
    5 & In \textit{Variant A}, it is 33.9\% more likely than in \textit{Variant B} that \taskize{IV Antibiotics} occurs in a process instance.\\\hline
    6 & In \textit{Variant A}, it is 33.9\% more likely than in \textit{Variant B} that if \taskize{ER Registration} occurs, also \taskize{IV Antibiotics} occurs.\\\hline
    7 & In \textit{Variant A}, it is 33.9\% more likely than in \textit{Variant B} that if \taskize{ER Sepsis Triage} occurs, \taskize{IV Antibiotics} will occur\\ & afterwards without any other occurrence of \taskize{ER Sepsis Triage} in between.\\\hline
    8 & In \textit{Variant A}, it is 33.9\% more likely than in \textit{Variant B} that if \taskize{ER Triage} occurs, also \taskize{IV Antibiotics} occurs.\\\hline
    \grayrow
    9 & In \textit{Variant A}, it is 31.2\% more likely than in \textit{Variant B} that \taskize{IV Liquid} occurs in a process instance.\\\hline
    10 & In \textit{Variant A}, it is 31.2\% more likely than in \textit{Variant B} that if \taskize{ER Registration} occurs, also \taskize{IV Liquid} occurs.\\\hline
    11 & In \textit{Variant A}, it is 31.2\% more likely than in \textit{Variant B} that if \taskize{ER Sepsis Triage} occurs, also \taskize{IV Liquid} occurs.\\\hline
    12 & In \textit{Variant A}, it is 31.2\% more likely than in \textit{Variant B} that if \taskize{ER Triage} occurs, also \taskize{IV Liquid} occurs.\\\hline
    
	\end{tabular}

	}
\end{table}
\begin{table}[tb]
	\caption{BPIC15 logs -- \gls{ourTechnique} output (our approach). 
		The list reports the first \num{10} top-ranked behavioural differences. } 
	\label{tab:bpi15}
	\centering	
	\resizebox{\textwidth}{!}{%
		\begin{tabular}{l|l}

    & \textbf{Statistically significant differences (\textit{Variant A}: $\mathrm{Municipality} = 1$; \textit{Variant B}: $\mathrm{Municipality}  = 2$)} \\\hline
	
	1 & It happens only in \textit{Variant A} that \taskize{01\_HOOFD\_456} may occur at most once in a process instance.\\\hline
    2 & It happens only in \textit{Variant A} that \taskize{01\_HOOFD\_492\_1} may occur at most once in a process instance.\\\hline
    3 & It happens only in \textit{Variant A} that \taskize{01\_HOOFD\_492\_2} may occur at most once in a process instance.\\\hline
    4 & It happens only in \textit{Variant A} that if \taskize{01\_HOOFD\_456} occurs, also \taskize{01\_HOOFD\_010} occurs.\\\hline
    5 & It happens only in \textit{Variant A} that if \taskize{01\_HOOFD\_456} occurs, also \taskize{01\_HOOFD\_015} occurs.\\\hline
    6 & It happens only in in \textit{Variant B} that if \taskize{16\_LGSD\_010} occurs, also \taskize{01\_HOOFD\_490\_2} occurs.\\\hline
    7 & It happens only in \textit{Variant B} that if \taskize{16\_LGSD\_010} occurs, also \taskize{01\_HOOFD\_495} occurs. \\\hline
    8 & It happens only in in \textit{Variant B} that if \taskize{16\_LGSD\_010} occurs, also \taskize{02\_DRZ\_010} occurs.\\\hline
    9 & It happens only in in \textit{Variant B} that if \taskize{16\_LGSD\_010} occurs, also \taskize{04\_BPT\_005} occurs.\\\hline
    10 & It happens only in in \textit{Variant B} that if \taskize{16\_LGSD\_010} occurs, also \taskize{09\_AH\_I\_010} occurs.\\\hline
    
\end{tabular}

	}
\end{table}

When the behaviour becomes less rigid and more flexible, the variants tend to have a higher percentage of distinct traces, like in the BPIC15 logs. In such cases, the output of MFVA becomes too simplistic (see \cref{fig:bpic15:1}) or so complex to the extent of being even barely interpretable (see \cref{fig:bpic15:2}). This result is, in general, typical of approaches based on graphs depicting processes imperatively~\cite{bolt2018process,nguyen2018multi,taymouri2020business}. In fact, the more flexible the behaviour recorded in the log, the larger the output graph (unless filtering is applied). On the other hand, our approach shows the differences as a list of declarative statements, focusing on the differences rather than on the overall picture 
(see \cref{tab:sepsis,tab:bpi15}). This is a benefit shared with PESVA, which also outputs natural language statements. However, PESVA suffers from scalability issues (PESVA timed out after running for three hours on BPIC15) and the differences identified by PESVA are limited in scope, which may hamper explanatory power and understandability. 
%
On the SEPSIS variants, e.g., PESVA indicates that ``Task \taskize{IV Liquid(4)} can be skipped in model 2, whereas in model 1 it is always executed''. Such a natural language formulation may be difficult to be interpreted, at times, in fact, task \taskize{IV Liquid(4)} does not refer to any execution of \taskize{IV Liquid}, but to the \taskize{IV Liquid} event that occurs as the fourth one. Our statements, instead, are broader in scope because they are not extrapolated by prime event structures but by declarative rules, therefore they refer to rules exerted on the whole process run rather than on the single occurrence of events.
Also, we remind that the differences captured by PESVA are not guaranteed to be statistically significant. 

In terms of execution time, \gls{ourTechnique} outperforms MFVA and PESVA when analysing the BPIC13, BPIC15, and SEPSIS logs:
considering the best execution times between those of MFVA and PESVA,
the runs required
\SI{81.8}{\second},
\SI{4901.3}{\second}, and
\SI{5.1}{\second}, respectively.
\Gls{ourTechnique} required instead
\SI{4.9}{\second},
\SI{326.7}{\second}, and
\SI{4.6}{\second}, respectively.
For the analysis of the RTFMP log, instead, our technique required
\SI{38.1}{\second}, thus outperforming MFVA
(\SI{1152.9}{\second}) but not PESVA
(\SI{21.7}{\second}).
Nevertheless, we underline that our approach appears to be more scalable than the MFVA and PESVA: to process the BPIC15 log, indeed, the former took more than an hour and a half, whereas the latter timed out at three hours.

%

\section{Conclusion}\label{sec:conclusion}
%
In this paper, we proposed a novel method to perform variant analysis and discover statistically-significant differences in the form of declarative rules expressed in natural language. 
We compared our method with state-of-the-art methods noticing that \gls{ourTechnique} provides a different level of expressiveness, easier output interpretation, and faster execution time.
Future research endeavours include the extension of our method to encompass the full spectrum of {\ltlpf} formulae~\cite{Cecconi2020ATemporal} and hybrid models~\cite{Slaats2020Declarative} as the rule language in place of standard {\Declare}, as well as multi-perspective specifications beyond the sole control-flow structure~\cite{schoenig2016multiperspective}. 
Also, we aim at further simplifying the output via redundancy-reduction techniques such as those in~\cite{DiCiccio2017Resolving}, to be adapted in order to improve the distinction of variants rather than reducing the rules of a single specifications. Other interesting avenues are the identification of statistically significant performance rules as opposed to behavioural rules, and the enhancement of the statistical soundness of the results addressing the multiple testing problem through {\pvalue} correction techniques~\cite{hamalainen2019tutorial}. 
%
\subsubsection{Acknowledgments.}
The work of C.\ Di Ciccio was partially funded by the Italian MIUR under grant ``Dipartimenti di eccellenza 2018-2022'' of the Department of Computer Science at Sapienza and by the Sapienza research project ``SPECTRA''.


\begin{thebibliography}{10}
\providecommand{\url}[1]{\texttt{#1}}
\providecommand{\urlprefix}{URL }
\providecommand{\doi}[1]{https://doi.org/#1}

\bibitem{VanDerAa2020Say}
van~der Aa, H., Balder, K.J., Maggi, F.M., Nolte, A.: Say it in your own words:
  Defining declarative process models using speech recognition. In: BPM Forum.
  pp. 51--67 (2020)

\bibitem{ProcessMiningBook2}
van~der Aalst, W.M.: Process {M}ining - {D}ata Science in Action. Springer
  (2016)

\bibitem{van2016log}
van Beest, N.R., Dumas, M., Garc{\'\i}a-Ba{\~n}uelos, L., La~Rosa, M.: Log
  delta analysis: Interpretable differencing of business process event logs.
  In: BPM. pp. 386--405 (2016)

\bibitem{bolt2018process}
Bolt, A., de~Leoni, M., van~der Aalst, W.M.: Process variant comparison: using
  event logs to detect differences in behavior and business rules. Information
  Systems  \textbf{74},  53--66 (2018)

\bibitem{Cecconi2020ATemporal}
Cecconi, A., {De Giacomo}, G., {Di Ciccio}, C., Maggi, F.M., Mendling, J.: A
  temporal logic-based measurement framework for process mining. In: ICPM. pp.
  113--120 (2020)

\bibitem{DiCiccio2017Resolving}
{Di Ciccio}, C., Maggi, F.M., Montali, M., Mendling, J.: Resolving
  inconsistencies and redundancies in declarative process models. Inf. Syst.
  \textbf{64},  425--446 (2017)

\bibitem{DBLP:journals/tmis/CiccioM15}
{Di Ciccio}, C., Mecella, M.: On the discovery of declarative control flows for
  artful processes. {ACM} Trans. Management Inf. Syst.  \textbf{5}(4),
  24:1--24:37 (2015)

\bibitem{edgington1969approximate}
Edgington, E.S.: Approximate randomization tests. The Journal of Psychology
  \textbf{72}(2),  143--149 (1969)

\bibitem{hamalainen2019tutorial}
H{\"a}m{\"a}l{\"a}inen, W., Webb, G.I.: A tutorial on statistically sound
  pattern discovery. Data Mining and Knowledge Discovery  \textbf{33}(2),
  325--377 (2019)

\bibitem{nguyen2018multi}
Nguyen, H., Dumas, M., La~Rosa, M., ter Hofstede, A.H.: Multi-perspective
  comparison of business process variants based on event logs. In: ER. pp.
  449--459 (2018)

\bibitem{nichols2002nonparametric}
Nichols, T.E., Holmes, A.P.: Nonparametric permutation tests for functional
  neuroimaging: a primer with examples. Human brain mapping  \textbf{15}(1),
  1--25 (2002)

\bibitem{pesic2010enacting}
Pesic, M., Bosnacki, D., van~der Aalst, W.M.: Enacting declarative languages
  using {LTL:} avoiding errors and improving performance. In: {SPIN}. pp.
  146--161 (2010)

\bibitem{pitman1937significance}
Pitman, E.J.: Significance tests which may be applied to samples from any
  populations. Supplement to the Journal of the Royal Statistical Society
  \textbf{4}(1),  119--130 (1937)

\bibitem{poelmans2010combining}
Poelmans, J., Dedene, G., Verheyden, G., Van~der Mussele, H., Viaene, S.,
  Peters, E.: Combining business process and data discovery techniques for
  analyzing and improving integrated care pathways. In: ICDM. pp. 505--517
  (2010)

\bibitem{schoenig2016multiperspective}
Sch{\"{o}}nig, S., {Di Ciccio}, C., Maggi, F.M., Mendling, J.: Discovery of
  multi-perspective declarative process models. In: {ICSOC}. pp. 87--103 (2016)

\bibitem{Slaats2020Declarative}
Slaats, T.: Declarative and hybrid process discovery: Recent advances and open
  challenges. J. Data Semant.  \textbf{9}(1),  3--20 (2020)

\bibitem{suriadi2013understanding}
Suriadi, S., Wynn, M.T., Ouyang, C., ter Hofstede, A.H., van Dijk, N.J.:
  Understanding process behaviours in a large insurance company in australia: A
  case study. In: CAiSE. pp. 449--464 (2013)

\bibitem{taymouri2020business}
Taymouri, F., La~Rosa, M., Carmona, J.: Business process variant analysis based
  on mutual fingerprints of event logs. In: CAiSE. pp. 299--318 (2020)

\bibitem{taymouri2021business}
Taymouri, F., La~Rosa, M., Dumas, M., Maggi, F.M.: Business process variant
  analysis: Survey and classification. Knowledge-Based Systems  \textbf{211},
  106557 (2021)

\bibitem{welch1990construction}
Welch, W.J.: Construction of permutation tests. Journal of the American
  Statistical Association  \textbf{85}(411),  693--698 (1990)

\bibitem{wu2016computing}
Wu, J., He, Z., Gu, F., Liu, X., Zhou, J., Yang, C.: Computing exact
  permutation p-values for association rules. Information Sciences
  \textbf{346},  146--162 (2016)

\end{thebibliography}

\end{document}